\documentclass[superscriptaddress,showpacs,showkeys,amsmath,amssymb,
a4paper,
aps,prl,preprint,
]{revtex4-1}

\usepackage{graphicx}
\usepackage{dcolumn}
\usepackage{bm}
\usepackage{hyperref}
\usepackage[svgnames]{xcolor}
\usepackage[colorinlistoftodos]{todonotes}

\begin{document}


\title{Large nonvolatile control of magnetic anisotropy in CoPt by a ferroelectric ZnO-based tunneling barrier}
\author{Muftah Al-Mahdawi}
\email{mahdawi@mlab.apph.tohoku.ac.jp}
\thanks{These authors had equal contributions to this work}
\affiliation{Department of Electronic Engineering, Tohoku University, Sendai 980-8579, Japan}
\affiliation{Center for Science and Innovation in Spintronics (Core Research Cluster), Tohoku University, Sendai 980-8577, Japan}
\affiliation{Center for Spintronics Research Network, Tohoku University, Sendai 980-8577, Japan}
\author{Mohamed Belmoubarik}
\thanks{These authors had equal contributions to this work}
\affiliation{Department of Electronic Engineering, Tohoku University, Sendai 980-8579, Japan}
\author{Masao Obata}
\email{obata@cphys.s.kanazawa-u.ac.jp}
\thanks{These authors had equal contributions to this work}
\affiliation{Graduate School of Natural Science and Technology, Kanazawa University, Kanazawa 920-1192, Japan}
\author{Daiki Yoshikawa}
\affiliation{Graduate School of Natural Science and Technology, Kanazawa University, Kanazawa 920-1192, Japan}
\author{Hideyuki Sato}
\affiliation{Department of Electronic Engineering, Tohoku University, Sendai 980-8579, Japan}
\author{Tomohiro Nozaki}
\affiliation{Department of Electronic Engineering, Tohoku University, Sendai 980-8579, Japan}
\author{Tatsuki Oda}
\affiliation{Graduate School of Natural Science and Technology, Kanazawa University, Kanazawa 920-1192, Japan}
\affiliation{Institute of Science and Engineering, Kanazawa University, Kanazawa 920-1192, Japan}
\author{Masashi Sahashi}
\affiliation{Department of Electronic Engineering, Tohoku University, Sendai 980-8579, Japan}
\affiliation{ImPACT Program, Japan Science and Technology Agency, Chiyoda, Tokyo 102-0076, Japan}

\date{\today}
\begin{abstract}
The electric control of magnetic anisotropy has important applications for nonvolatile memory and information processing. By first-principles calculations, we show a large nonvolatile control of magnetic anisotropy in ferromagnetic/ferroelectric CoPt/ZnO interface. Using the switched electric polarization of ZnO, the density-of-states and magnetic anisotropy at the CoPt surface show a large change. Due to a strong Co/Pt orbitals hybridization and a large spin-orbit coupling, a large control of magnetic anisotropy was found. We experimentally measured the change of effective anisotropy by tunneling resistance measurements in CoPt/Mg-doped ZnO/Co junctions. Additionally, we corroborate the origin of the control of magnetic anisotropy by observations on tunneling anisotropic magnetoresistance.
\end{abstract}

\maketitle
\paragraph{Introduction}---
The recent strong interest in the control of magnetism by electric means has been driven by the aspects of fundamental physical understanding of magnetism, and more importantly towards applications in nonvolatile information processing in magnetic memories \cite{matsukura_2015,song_2017}. The electric field effect on the interfacial electronic states has been reported to control magnetic Curie temperature $T_C$ \cite{ohno_2000,chiba_2011}, coercivity $H_C$ \cite{chiba_2003,weisheit_2007,endo_2010}, magnetic moment \cite{obinata_2015}, spin polarization \cite{garcia_2010}, and magnetic anisotropy energy MAE \cite{chiba_2008,maruyama_2009} both in magnetic semiconductors and transition metals. By using the voltage control of MAE (VCMA), fast writing by nanosecond electric pulses in magnetic tunnel junctions (MTJs) has been widely demonstrated in MTJs based on rock-salt-type MgO barrier \cite{shiota_2011,wang_2012,kanai_2012}. The main electronic origin of interfacial MAE, and hence VCMA, in an Fe-alloy/MgO system is the orbital hybridization between Fe $3d$ and O $2p$ at the interface in the presence of spin-orbit coupling (SOC) \cite{yang2011}. However, the precise control of Fe surface oxidation is crucial \cite{yang2011} for large MAE and VCMA magnitudes, which are required for nonvolatile memory applications \cite{amiri_2015,grezes_2017,shiota_2017}.

An alternative is the heterostructure of a ferromagnet/ferroelectric (FM/FE) combination \cite{duan_2008a,mardana_2011,lukashev2012,lukashev_2012a,lee_2013b}, where a relatively large modulation of MAE is achieved by the control of FE polarization ($P$). This was shown to be mainly due to the $P$-dependent hybridization between the orbitals of the FM and FE elements. 
In this work, we propose and clarify an FM/FE MTJ system for achieving a large nonvolatile control of MAE, with a different mechanism. We investigated MTJs made from fcc-CoPt FM and wurtzite-ZnO FE, using first-principles calculations and magnetotransport measurements.

An important aspect of $3d$-$5d$ magnetic alloys is that MAE originates from the large SOC of the $5d$ element which is magnetized by the strong exchange field of $3d$ moments \cite{grange_1998a,mryasov_2005,shick_2008}. In CoPt \cite{sakuma_1994,grange_1998a}, a large MAE along the $(1\,1\,1)$ crystal axis is retained even under disorder \cite{weller_1992}. Wurtzite-type ZnO has a direct bandgap of 3.2 eV, and it becomes FE by Mg doping with enhanced insulating property \cite{kozuka2014}. The electric dipoles of ZnO are aligned along $(0\,0\,0\,1)$ direction. The $P$ reversal changes the chemical potential at interface, that should have a prominent effect on MAE due to the large $5d$ SOC \cite{shick_2003}. Therefore, the system of CoPt $(1\,1\,1)$/Mg-ZnO $(0\,0\,0\,1)$\cite{belmoubarik2015,belmoubarik2016} has a strong potential for a large control of MAE.

\begin{figure}
	\caption{(a) A schematic of the MgZnO magnetic tunnel junction (MTJ) and the electric-field cooling (EFC) procedure. (b) The reversal of MgZnO electric polarization ($P$) after EFC changes the surface charge at the metal interfaces of the MTJ. The charge difference is negative (positive) at the CoPt surface after $+$EFC ($-$EFC). Correspondingly, the surface MAE of CoPt is perpendicular (inplane) for $P^+$ ($P^-$) states. (c) Schematics of the MTJ potential profile at $P^\pm$ states.}
	\label{fig:schem}
	\includegraphics[width=0.7\textwidth]{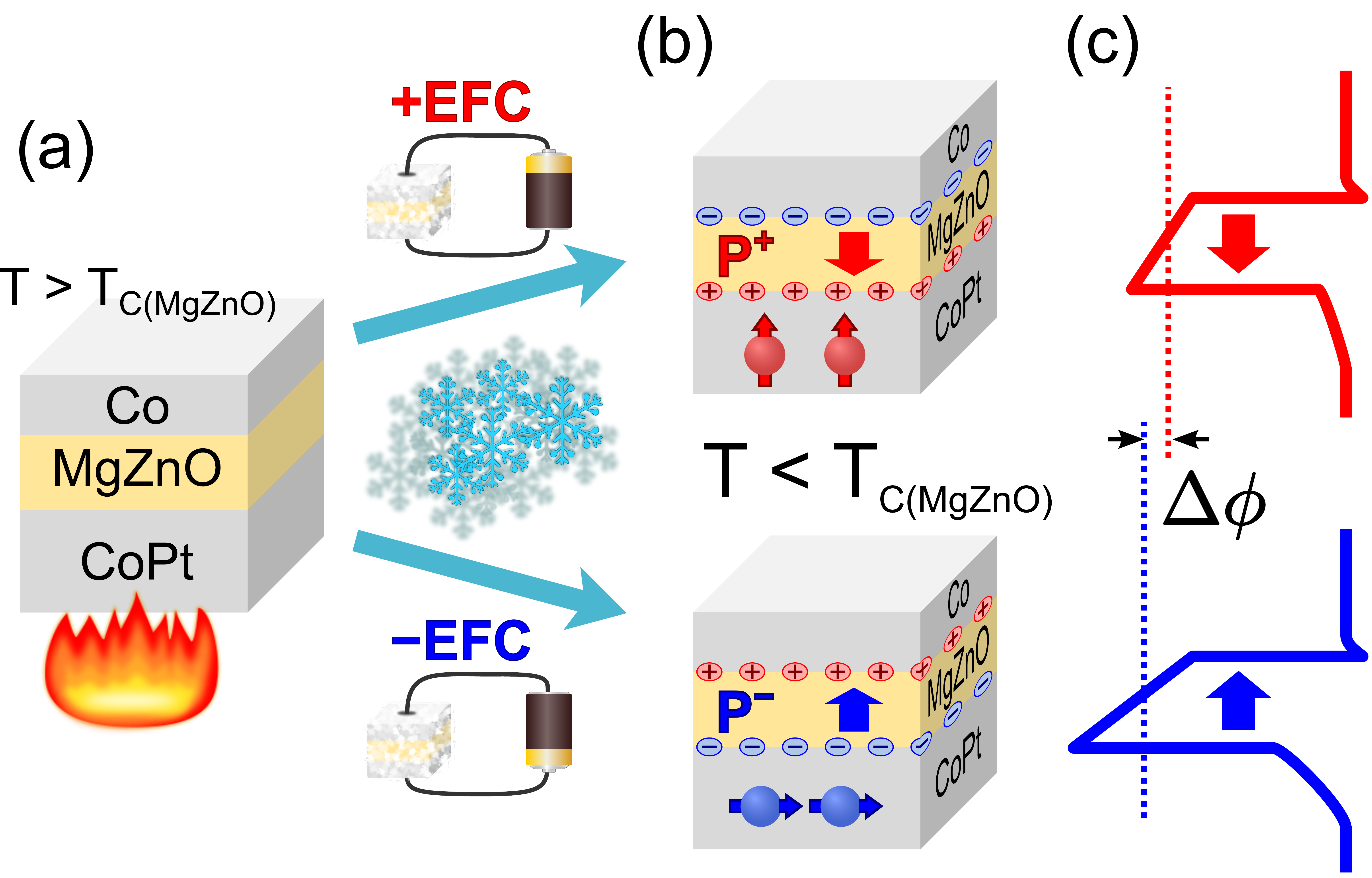}
\end{figure}

The schematic of the device and the effect of $P$ on MAE are shown in Fig.~\ref{fig:schem}, together with the experimental procedure employed for this report. Due to the large electric coercivity of MgZnO, the cooling under an electric field (EFC) from above the MgZnO Curie temperature was used to align $P$ in either direction [Fig.~\ref{fig:schem}(a)] \cite{belmoubarik2016}. The reversal of $P$ results in a difference of the equilibrium surface charge at the surfaces of the FM electrodes [Fig.~\ref{fig:schem}(b)]. The charge difference is negative (positive) at the CoPt surface after $+$EFC ($-$EFC), due to the formation of a positive (negative) $P$ state [Fig.~\ref{fig:schem}(b,c)]. Based on the following theoretical and experimental analyses, the surface MAE of CoPt electrode will be perpendicular (inplane) for a $P^+$ ($P^-$) state. Such a large non-volatile change of MAE is driven by the $P$-modulation of the SOC and the $3d$-$5d$ hybridization.

\begin{figure}
	\caption{(a,b) The model of CoPt/ZnO used for the first-principles calculations, in the (a) $P^+$ and (b) $P^-$ states. (c--f) The projected density of states (PDOS) the (c,d) Co and (e,f) Pt-1 atoms in the the (c,e) $P^+$ and (d,f) $P^-$ states.}
	\label{fig:sim}
	\includegraphics[width=0.8\textwidth]{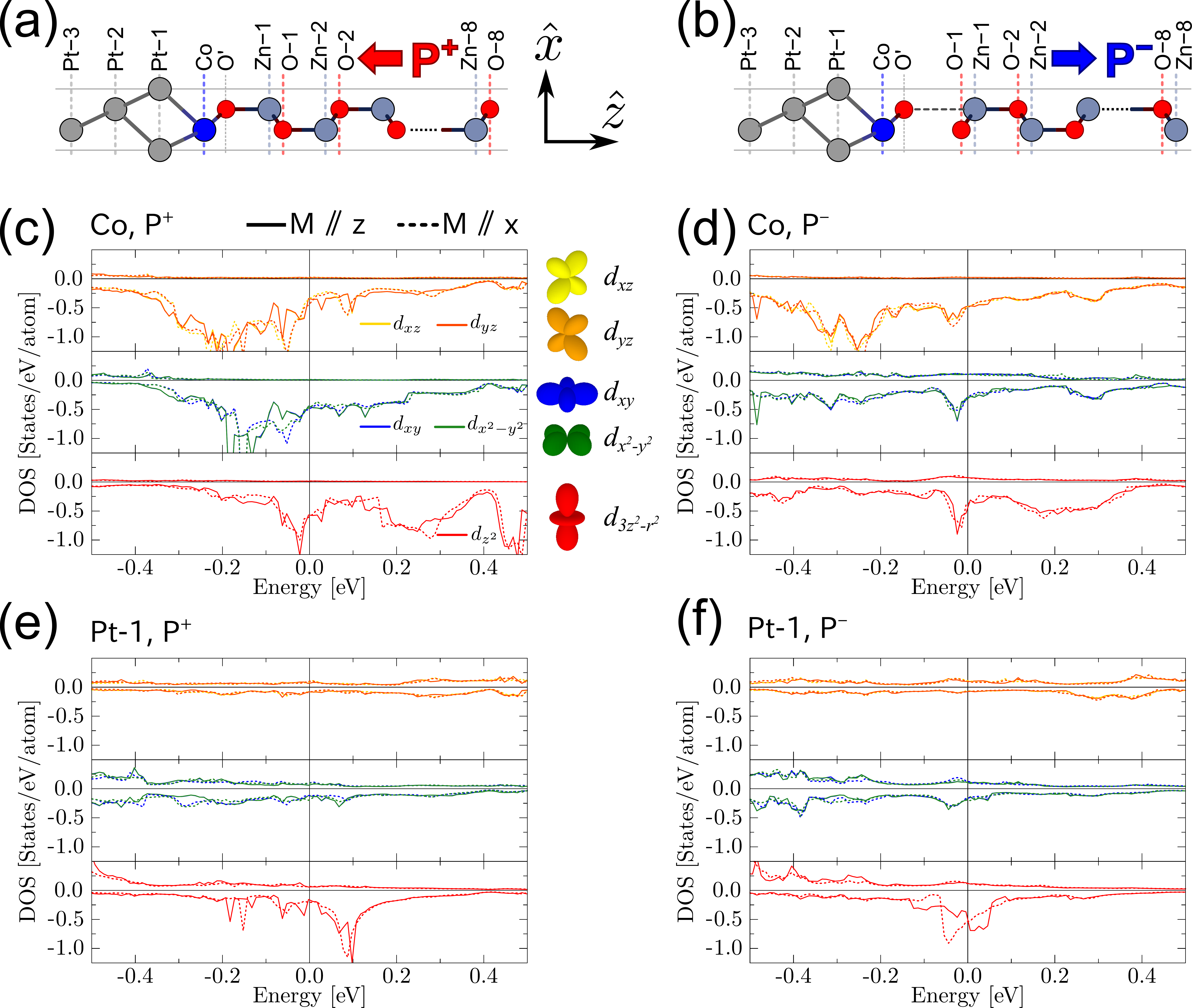}
\end{figure}

\paragraph{First-principles calculations}---
We modeled the effect of ZnO electric polarization on the interfacial MAE of CoPt by first-principles density functional theory calculations \cite{hohenberg1964}. We used scalar and fully relativistic ultra-soft pseudopotentials (USPPs) \cite{vanderbilt1990,tsujikawa_2008} and a plane-wave basis with the generalized gradient approximation \cite{perdew1992}. A change from a 24$\times$24$\times$1 mesh to a 32$\times$32$\times$1 mesh in the $k$-point sampling space did not show a significant difference in results, and the 32$\times$32$\times$1 mesh was used. The energy cut-off was set at 30 (300) Ry for the planewave basis in wavefunction (electron density). The structure was vacuum/Pt(3)/Co(1)/O(1)/Zn-O(8)/vacuum (atomic monolayers or bilayers) as shown in Figs.~\ref{fig:sim}(a,b). The $P^+$ and $P^-$ states were modeled as either Zn-O or O-Zn bilayers, respectively. The considerations behind the choices for the modeled structure and more details of the calculation methods are available in Ref.~\cite{belmoubarik2016}. The density of states (DOS) was calculated for magnetization orientations at the inplane $\hat{x}$ and out-of-plane $\hat{z}$ directions for each of $P^+$ and $P^-$ states, while the SOC was included [Figs.~\ref{fig:sim}(c--f)]. In the following, we define the changes of relevant quantities with respect to $P^+$ state $[\Delta w \!=\! w(P^-) \!-\! w(P^+)]$, whereas the anisotropy with respect to inplane direction [$\delta w \!=\! w(M\!\!\parallel\!\!z) - w(M\!\!\parallel\!\! x)$]. At each $P$ state, the MAE was calculated as the difference between the total energies at the inplane and out-of-plane magnetization $M$ configurations, \emph{i.e.}~$\mathrm{MAE} \!=\! E_x \!-\! E_z$. A positive (negative) sign denotes an out-of-plane (inplane) MAE.

\begin{figure}
	\caption{The layer-resolved values of (a) the change of number of electrons ($\Delta n$), (b) the spin momentum ($m_s$), its change by $P$ ($\Delta m_s$), and anisotropy ($\delta m_s$), (c) the orbital momentum ($m_o$), its change by $P$ ($\Delta m_o$), and anisotropy ($\delta m_o$).
    }
	\label{fig:layers}
	\includegraphics[width=0.8\textwidth]{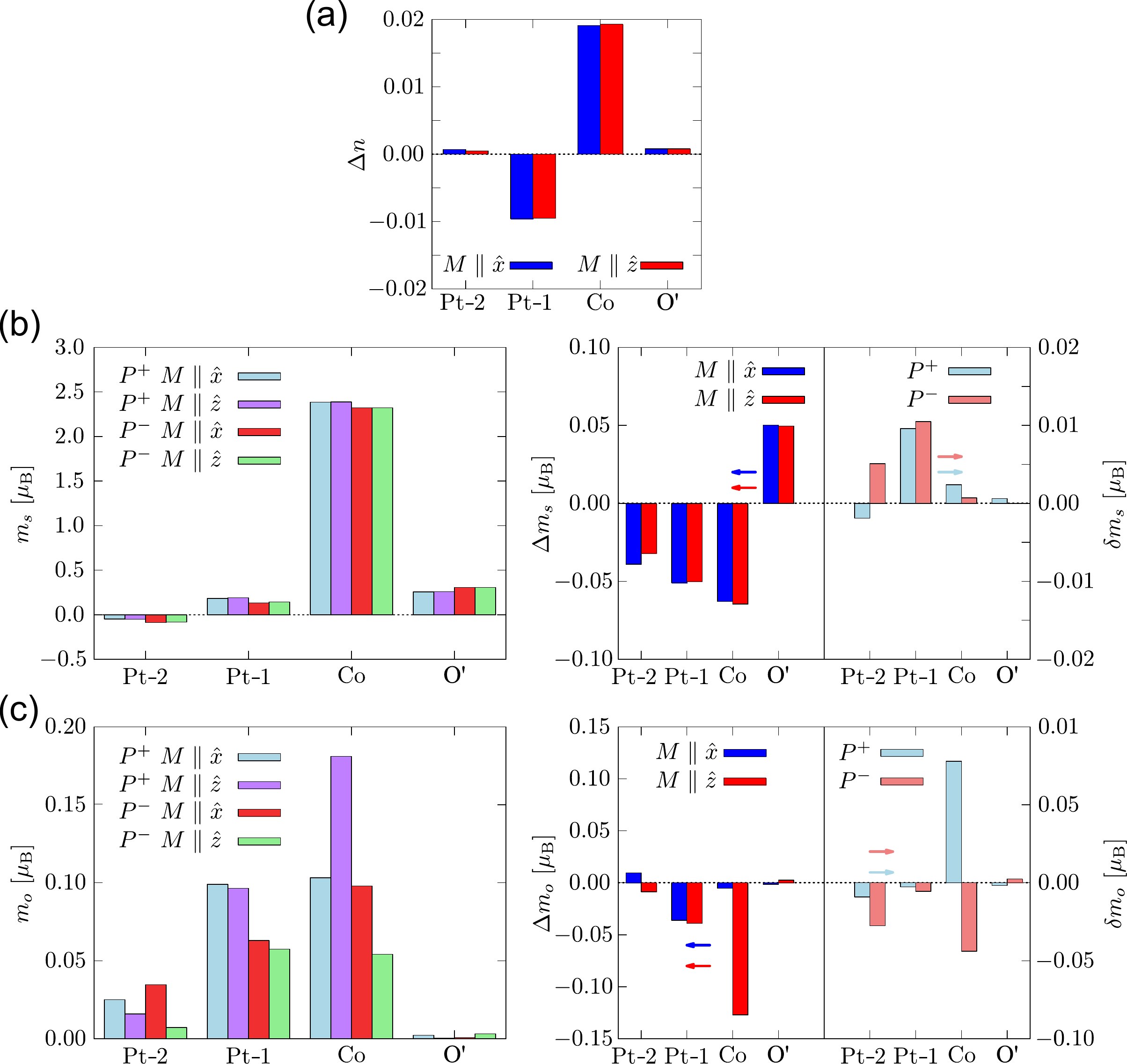}
\end{figure}

A drastic change of DOS was found between $P^+$ and $P^-$ states [Figs.~\ref{fig:sim}(c--f)]. The change of number of electrons ($\Delta n$) at the interface Co atom is $+$0.02 [Fig. \ref{fig:layers}(a)], where the $P^+$ ($P^-$) state corresponds to an electron depletion from (doping to) Pt-Co-O interface atoms, as depicted in Fig.~\ref{fig:schem}(b). Surface MAE was $+$0.25 and $-$1.23 meV/atom for $P^+$ and $P^-$ states, respectively. These correspond to $+$0.44 and $-$2.16 erg/cm$^2$ ($\equiv \mathrm{mJ/m^2}$). The electrons depletion increased the out-of-plane MAE, same as the interfaces of Fe/MgO \cite{maruyama2009,nakamura2010,niranjan2010,yoshikawa2014,miwa2015,nozaki2016,qingyi2017} and (Fe,Co)/Pt/MgO \cite{zhang2009,tsujikawa2012,miwa2017}. Although $\Delta n$ is close in value to an Fe/MgO interface under an electric field of 0.5--1.0 V/nm, the present $\Delta\mathrm{MAE}$ of $-$2.60 erg/cm$^2$ in CoPt/ZnO is much larger than Fe/MgO, which ranges at $-$0.1 -- $-$0.2 erg/cm$^2$ \cite{nakamura2010,niranjan2010,yoshikawa2014}. Among the 3\emph{d}-noble metal alloys, calculations showed that CoPt-based systems have the largest electric-field effect on MAE \cite{zhang2009,tsujikawa2012}. We need to emphasize that $\Delta n$ in the presented CoPt/ZnO system comes from the electric polarization of ZnO, without the need of a continuously applied voltage as in Fe/MgO.

This large $\Delta\mathrm{MAE} / \Delta n$ can be explained by the control of $3d$-$5d$ orbital hybridization of Co/Pt. 
There is a hybridized peak in $d_{3z^2-r^2} (m=0)$ minority spin DOS, that moves near the Fermi level ($E_\mathrm{F}$) with the change of $P$ direction [Fig.~\ref{fig:sim}(e,f)].  Simultaneously, due to the electric field change, the part of electrons occupying the $3d$ orbitals extending to $xy$-plane directions ($d_{xz,yz}$ for $|m|\!=\!1$ and $d_{xy,x^{2}\!-\!y^{2}}$ for $|m|\!=\!2$) are redistributed into those of $d_{3z^2-r^{2}}$. This produces a change in orbital and spin momenta of Co and Pt, leading to a large change in DOS and MAE \cite{daalderop_1990,daalderop_1994,sakuma_1994}.
Moreover, while the SOC of Co, which is enhanced by Pt, is favoring a perpendicular MAE, the $P$-modulation resulted in the closer participation of Pt into the electronic structure at $E_\mathrm{F}$ [Fig.~\ref{fig:sim}(f)]. This induces large reductions of orbital momenta on both Co and Pt [Fig. \ref{fig:layers}(c)], leading to an additional decrease in MAE, due to the strong SOC of Pt. Therefore, even a small $\Delta n$ in the CoPt/ZnO system gives a strong modulation of MAE. 
 
The DOS sets of $d_{xz,yz}$ and $d_{xy,x^2-y^2}$ orbitals are degenerate in the $M\!\!\parallel\!\!\hat{z}$ direction, whereas the degeneracies are lifted when $M\!\!\parallel\!\!\hat{x}$ [Fig.~\ref{fig:sim}(c--f)]. At $P^+$, based on Bruno's model relating to down-down spin scattering \cite{bruno_1989}, both sets and the large SOC contribute strongly to a perpendicular MAE \cite{wang_1993a}. Correspondingly, the anisotropy in orbital momentum ($\delta m_o$) of Co is large at {$+0.078 \mu_B$}, supporting the origin of a perpendicular MAE [Fig.~\ref{fig:layers}(c)]. At $P^-$ state, the previously-unoccupied Pt $d_{3z^2-r^2}$ minority peak in DOS is shifted towards $E_\mathrm{F}$, which increases the density of minority spins [Fig.~\ref{fig:sim}(f)]. The application of Bruno's model is limited by the presence of majority states near $E_\mathrm{F}$ for both of Co and Pt, and an anisotropy in spin momentum ({$\delta m_s = +0.010 \mu_B$}) of Pt-1, due to the hybridization change depending on the orientation of $M$ [Fig.~\ref{fig:layers}(c)]. However, $\delta m_o$ of Co is negative at {$-0.044 \mu_B$}, indicating an inplane MAE.

\begin{figure}
	\caption{The experimental effect of $P$ on MAE. (a) The $M$--$H$ curves in the inplane and out-of-plane directions of the as-deposited state ($P^+$) of blanket films. The CoPt layer shows a small inplane total MAE, whereas the Co layer has a dominant inplane shape MAE. (b,c) The $R$--$H$ curves of a fabricated junction in the inplane and perpendicular directions (b) after $+$EFC at $P^+$ state, and (c) after $-$EFC at $P^-$ state. The $R$--$H$ curves indicate that the surface MAE of CoPt changed from out-of-plane to inplane direction by $P$-modulation. The measurements in (a) and (b,c) were done at 5 K and 2 K, respectively.}
	\label{fig:RH}
	\includegraphics[width=0.5\textwidth]{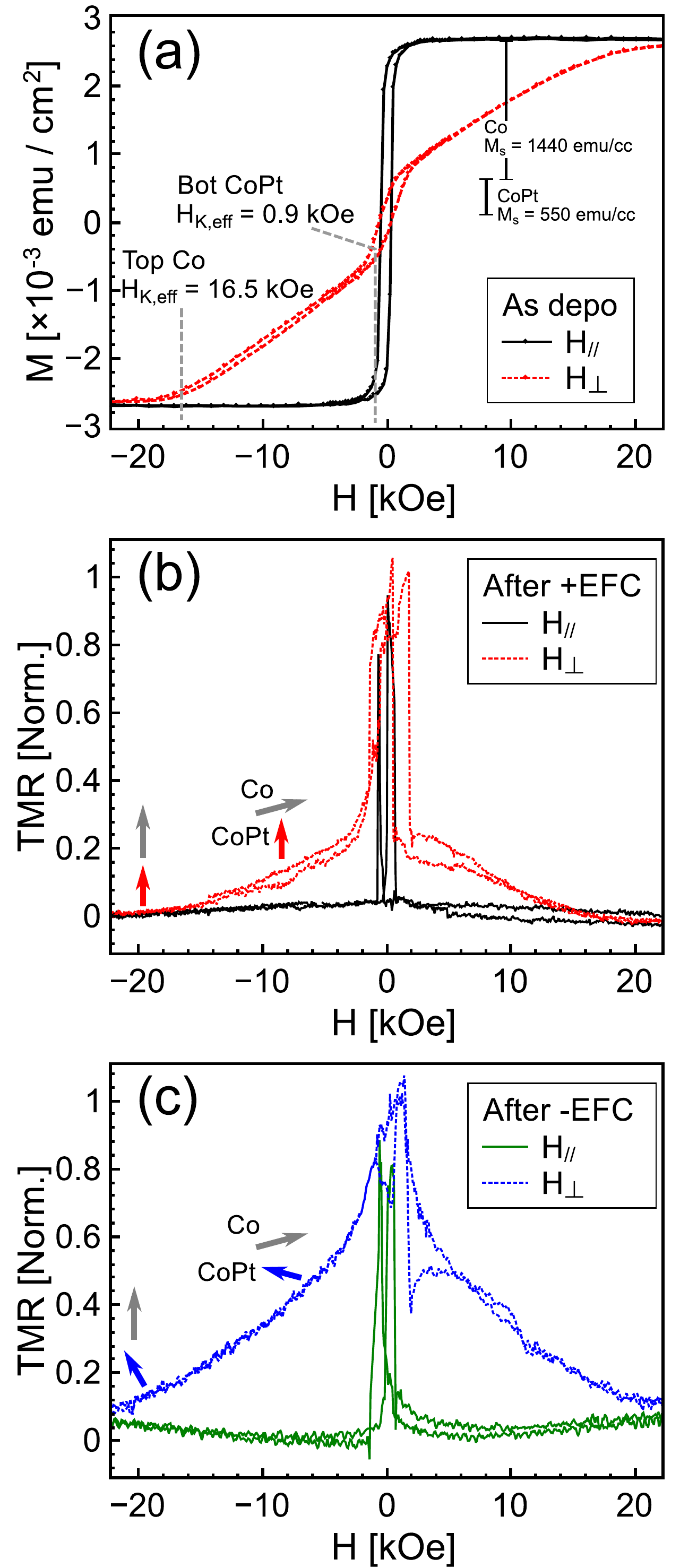}
\end{figure}

\paragraph{MAE observations from resistance-field curves}---
We conducted experimental confirmations, using observations by resistance-field ($R$--$H$) curves with the field applied in the $\hat{x}$ and $\hat{z}$ directions. The presented experimental results were obtained from samples described previously \cite{belmoubarik2015,belmoubarik2016}. The initial film structure was made of (thicknesses in nm): \emph{c}-plane sapphire Al$_2$O$_3$ substrate/Pt (30)/Co$_{0.3}$Pt$_{0.7}$ (10)/Mg$_{0.23}$Zn$_{0.77}$ (7)/Co (16). The unpatterned films were used for magnetic and micro-structure characterization. The metallic layers were grown by sputtering and electron-beam evaporation, whereas MgZnO was grown from Mg and Zn metallic sources using molecular-beam epitaxy equipped with an oxygen radical source. The structural analysis showed an epitaxial growth of fcc-Pt/fcc-CoPt/w-MgZnO/hcp-Co. The $(1\,1\,1)$-oriented growth of Pt allows for the growth of MgZnO along the polar c-axis, which is suitable to the control of metal surface charge. A relatively-thick bottom electrode was needed for better growth of the MgZnO barrier. Therefore, the deposition conditions and composition of CoPt were chosen to have a near compensation to the shape anisotropy by the bulk perpendicular MAE. At such a compensation, the changes in anisotropy at CoPt/MgZnO interface will be observable, even with a rather thick CoPt layer. We measured the magnetic properties by a Superconducting Quantum Interference Device (SQUID) magnetometer of an unfabricated film in the as-deposited state, which corresponds to $P^+$. The inplane and out-of-plane $M$--$H$ curves measured at 5 K show that the top Co and bottom CoPt layers have an inplane easy axis at $P^+$ state [Fig.~\ref{fig:RH}(a)]. However, the bottom CoPt has a small perpendicular saturation field ($H_\mathrm{K,eff} = 0.9$ kOe), due to the compensation mentioned earlier.

For the fabrication of MTJs and to induce coercivity difference between the top and bottom ferromagnetic layers, the top Co layer was etched down to 2 nm, then Co$_{0.5}$Fe$_{0.5}$ (2)/IrMn (14)/Ru (5) was deposited \emph{in situ}. After that, the stack was pin-annealed for 30 min at 270$^\circ$C and a 10-kOe magnetic field. The presented tunneling resistance results are from circular junctions 10 $\mu$m in diameter, microfabricated by electron-beam lithography and Ar-ion milling. The $R$--$H$ curves were measured at 2 K, after using the EFC procedure to align the MgZnO $P$ \cite{belmoubarik2016}. The $\pm$EFC from 360 K down to 2 K corresponds to the $P^\pm$ states. The normalized $R$--$H$ curves after $\pm$EFC are shown in Figs.~\ref{fig:RH}(b,c). At the $P^+$ state, the $R$--$H$ curves show similar character to the as-deposited $M$--$H$ curves. In the out-of-plane field direction, there are two shoulders at 1 kOe and 16 kOe, corresponding to $H_\mathrm{K,eff}$ of CoPt and Co, respectively. Above the saturation of CoPt at 1 kOe, the CoPt magnetization is out-of-plane, and the Co magnetization rotates towards the out-of-plane direction until saturation at 16 kOe [arrows in Fig.~\ref{fig:RH}(b)]. On the other hand, at $P^-$ state a drastic change in CoPt anisotropy happens. The area enclosed by inplane and out-of-plane curves is much larger, and a field of 20 kOe is required to saturate the resistance [Fig.~\ref{fig:RH}(c)]. This can be explained by a change of CoPt easy-axis to the inplane direction, with $H_\mathrm{K,eff}$ = 20 kOe [arrows in Fig.~\ref{fig:RH}(c)]. The areal MAE of CoPt ($K_s$) can be found from the following relation:
\begin{equation}
K_s = (4 \pi M_s - H_\mathrm{K,eff} ) \times \frac{M_s t}{2},
\end{equation}
where $M_s$ is the saturation magnetization, and $t$ is CoPt thickness. The corresponding results are $K_s$ = $+1.6$ and $-3.6$ erg/cm$^2$ for $P^+$ and $P^-$ states, respectively. The aforementioned theoretical value of $\Delta \mathrm{MAE}$ = $-2.6$ erg/cm$^2$ has a good agreement with the experimental $\Delta \mathrm{MAE}$ = $-5.2$ erg/cm$^2$. The direction of change is same, and the magnitude is within a factor of 2.

\paragraph{Tunneling anisotropic magnetoresistance}---
A consequence of an anisotropic DOS (ADOS) is the dependence of tunneling current on $M$ direction, the named tunneling anisotropic magnetoresistance (TAMR) \cite{gould_2004,park_2008}. As the TAMR and MAE in $3d$-$5d$ systems come from the same origin, the study of TAMR can put light on MAE \cite{saito_2005}, and can confirm that the first-principle calculations correlate with the experiments on MAE. In the present case, the band-selective filtering of tunneling current is not evident, and TAMR in CoPt-based tunnel junctions can be understood qualitatively in terms of ADOS at the interface next to the tunneling barrier \cite{shick_2006,park_2008}. Therefore, we treat TAMR as originating from the interfacial ADOS in CoPt. 

\begin{figure}
	\caption{The effect of $P$-modulation on tunneling anisotropic magnetoresistance (TAMR). (a,b) Samples of the dependence of TAMR on applied field angle ($\phi$) at different bias voltages, (a) after $+$EFC at $P^+$ state, and (b) after $-$EFC at $P^-$ state. The curves are vertically shifted uniformly for clarity. (c) The experimental bias dependence of TAMR. (d) The energy dependence of TAMR from the first-principles calculations. A qualitative agreement is found between the experimental and calculation results.}
	\label{fig:TAMR}
	\includegraphics[width=0.8\textwidth]{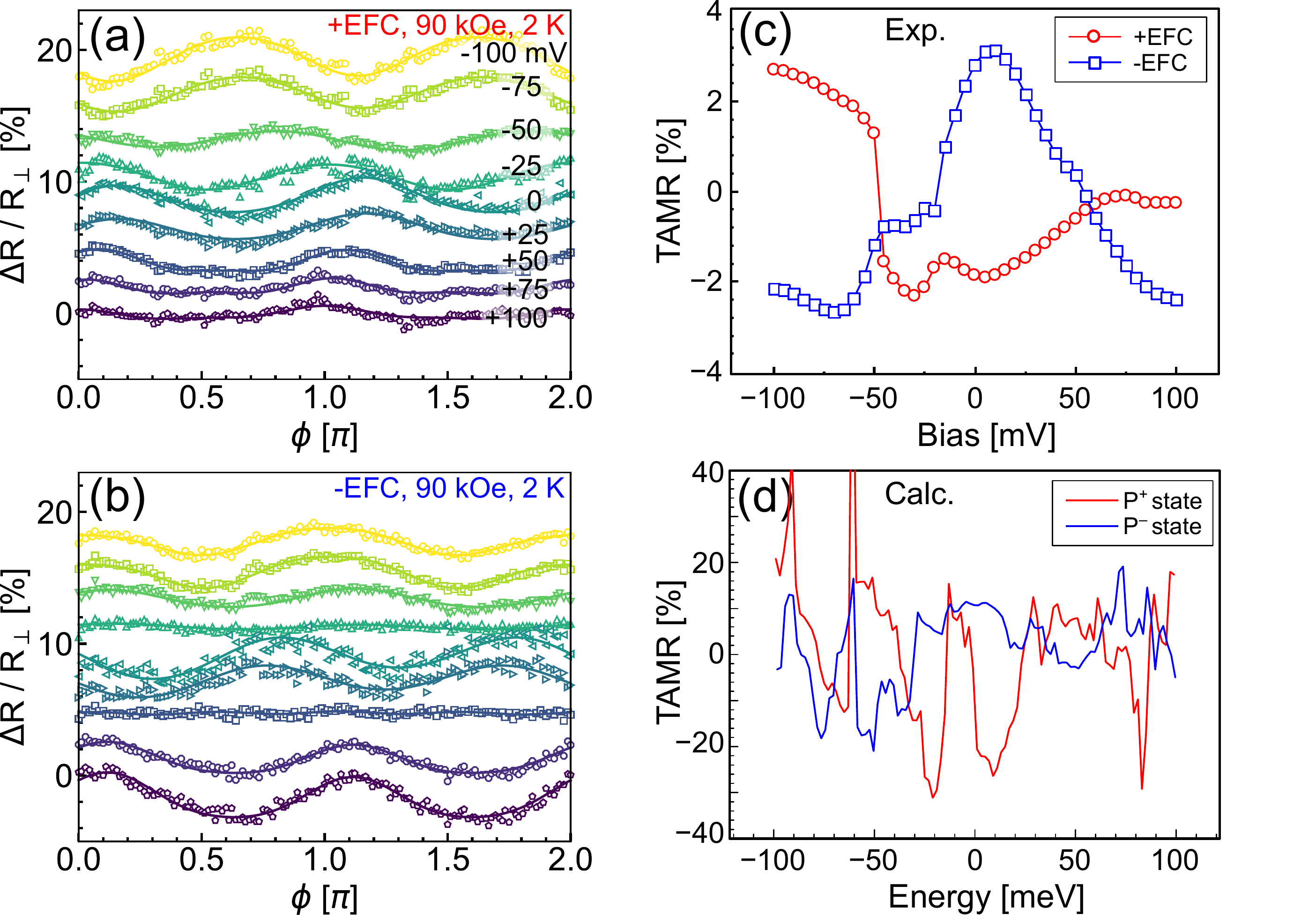}
\end{figure}

In the calculations, the direction of ADOS changed sign between the two $P$ states [Fig.~\ref{fig:sim}]. The anisotropy is mainly in the minority spin, and showed a character similar to MAE mentioned earlier. For $P^+$ ($P^-$) state, ADOS is mainly at $d_{xz,yz}$ and $d_{xy,x^2-y^2}$ ($d_{3z^2-r^2}$) orbitals. The total ADOS, defined as $\mathrm{DOS}(M\!\!\parallel\!\! z)/ \mathrm{DOS}(M\!\!\parallel\!\! x) \!-\! 1$, shows at $E_\mathrm{F}$ a sign change from negative at $P^+$ state to positive at $P^-$ [Fig.~\ref{fig:TAMR}(d)]. Experimentally, TAMR was measured from the out-of-plane angular dependence of differential resistance ($R$-$\phi$) at 90 kOe, which is much higher than saturation. Figs.~\ref{fig:TAMR}(a,b) show the normalized $R$-$\phi$ curves taken at various biases. TAMR was extracted from the twofold component and shown in Fig.\ref{fig:TAMR}(c). A qualitative agreement between experiments and calculations in TAMR sign and energy dependence is found. Therefore, our first-principles calculations captured the origin of a large $\Delta \mathrm{MAE}$ in CoPt/ZnO hexagonal system.

\paragraph{Conclusions}---
In summary, by first-principles calculations and experimentally, we investigated the large control of MAE of CoPt ferromagnet by the electric polarization of wurtzite MgZnO tunnel barrier. The surface MAE sign changed, with a difference in magnitude that is much larger than Fe/MgO. A combined study of MAE and TAMR showed consistent results from both experiments and calculations. Therefore, we consider that the TAMR observations are important for the explanation of $\Delta \mathrm{MAE}$. The origin is likely due to the control of DOS and SOC in CoPt interface by the modulation of $3d$-$5d$ hybridization induced by ZnO polarization. This shows the possibility of designing large MAE.

As a final note on possible applications, the ZnO polarization can be used either as an amplifier for low-voltage control of MAE, or for non-volatile gating of MAE.  One possibility is in the toggle-type voltage magnetic random-access memories (V-MRAM) \cite{shiota_2011,kanai_2012}. An alternating $+ - +$ voltage pulse can be used for the precessional magnetization switching, similar to what is proposed for non-FE barriers \cite{ikeura_2018}. In the VCMA-assisted spin-orbit-torque writing \cite{yoda_2016}, we propose that a single voltage pulse can be used for non-volatile bit selection/deactivation in Pt/CoPt/ZnO structures. The other bits do not require manipulation, therefore making the scheme much simpler. Another favorable application is the utilization of non-volatile control of MAE in reconfigurable spinwave logic devices \cite{rana_2018}. Therefore, we believe that the present work should open a way for applications in nonvolatile energy-efficient control of magnetic memories and information processing.

\begin{acknowledgments}
The authors are thankful to Haruyuki Endo of Iwate Industrial Research Institute, Japan for helping in the samples preparation.
This work was partially supported by the JSPS Grant Number JP13J05806, 
the JSPS KAKENHI Grant Number JP18K04923, 
Kanazawa University SAKIGAKE Project, 
the Computational Materials Science Initiative, Japan, 
the Advanced Institute for Computational Science and Information Technology Center of Nagoya University through the High Performance Computing Infrastructure (HPCI) System Research Project under Projects hp17168 and hp180206, 
and the ImPACT Program of the Council for Science, Technology and Innovation (Cabinet Office, Government of Japan). The first-principles calculations were performed using the facilities of the Supercomputer Center, Institute for Solid State Physics, University of Tokyo, Japan. 
\end{acknowledgments}

\clearpage
\newpage
\bibliographystyle{apsrev4-1}
\bibliography{refs}

\begin{thebibliography}{57}%
\makeatletter
\providecommand \@ifxundefined [1]{%
 \@ifx{#1\undefined}
}%
\providecommand \@ifnum [1]{%
 \ifnum #1\expandafter \@firstoftwo
 \else \expandafter \@secondoftwo
 \fi
}%
\providecommand \@ifx [1]{%
 \ifx #1\expandafter \@firstoftwo
 \else \expandafter \@secondoftwo
 \fi
}%
\providecommand \natexlab [1]{#1}%
\providecommand \enquote  [1]{``#1''}%
\providecommand \bibnamefont  [1]{#1}%
\providecommand \bibfnamefont [1]{#1}%
\providecommand \citenamefont [1]{#1}%
\providecommand \href@noop [0]{\@secondoftwo}%
\providecommand \href [0]{\begingroup \@sanitize@url \@href}%
\providecommand \@href[1]{\@@startlink{#1}\@@href}%
\providecommand \@@href[1]{\endgroup#1\@@endlink}%
\providecommand \@sanitize@url [0]{\catcode `\\12\catcode `\$12\catcode
  `\&12\catcode `\#12\catcode `\^12\catcode `\_12\catcode `\%12\relax}%
\providecommand \@@startlink[1]{}%
\providecommand \@@endlink[0]{}%
\providecommand \url  [0]{\begingroup\@sanitize@url \@url }%
\providecommand \@url [1]{\endgroup\@href {#1}{\urlprefix }}%
\providecommand \urlprefix  [0]{URL }%
\providecommand \Eprint [0]{\href }%
\providecommand \doibase [0]{http://dx.doi.org/}%
\providecommand \selectlanguage [0]{\@gobble}%
\providecommand \bibinfo  [0]{\@secondoftwo}%
\providecommand \bibfield  [0]{\@secondoftwo}%
\providecommand \translation [1]{[#1]}%
\providecommand \BibitemOpen [0]{}%
\providecommand \bibitemStop [0]{}%
\providecommand \bibitemNoStop [0]{.\EOS\space}%
\providecommand \EOS [0]{\spacefactor3000\relax}%
\providecommand \BibitemShut  [1]{\csname bibitem#1\endcsname}%
\let\auto@bib@innerbib\@empty
\bibitem [{\citenamefont {Matsukura}\ \emph {et~al.}(2015)\citenamefont
  {Matsukura}, \citenamefont {Tokura},\ and\ \citenamefont
  {Ohno}}]{matsukura_2015}%
  \BibitemOpen
  \bibfield  {author} {\bibinfo {author} {\bibfnamefont {F.}~\bibnamefont
  {Matsukura}}, \bibinfo {author} {\bibfnamefont {Y.}~\bibnamefont {Tokura}}, \
  and\ \bibinfo {author} {\bibfnamefont {H.}~\bibnamefont {Ohno}},\ }\href
  {\doibase 10.1038/nnano.2015.22} {\bibfield  {journal} {\bibinfo  {journal}
  {Nature Nanotechnology}\ }\textbf {\bibinfo {volume} {10}},\ \bibinfo {pages}
  {209} (\bibinfo {year} {2015})}\BibitemShut {NoStop}%
\bibitem [{\citenamefont {Song}\ \emph {et~al.}(2017)\citenamefont {Song},
  \citenamefont {Cui}, \citenamefont {Li}, \citenamefont {Zhou},\ and\
  \citenamefont {Pan}}]{song_2017}%
  \BibitemOpen
  \bibfield  {author} {\bibinfo {author} {\bibfnamefont {C.}~\bibnamefont
  {Song}}, \bibinfo {author} {\bibfnamefont {B.}~\bibnamefont {Cui}}, \bibinfo
  {author} {\bibfnamefont {F.}~\bibnamefont {Li}}, \bibinfo {author}
  {\bibfnamefont {X.}~\bibnamefont {Zhou}}, \ and\ \bibinfo {author}
  {\bibfnamefont {F.}~\bibnamefont {Pan}},\ }\href {\doibase
  10.1016/j.pmatsci.2017.02.002} {\bibfield  {journal} {\bibinfo  {journal}
  {Progress in Materials Science}\ }\textbf {\bibinfo {volume} {87}},\ \bibinfo
  {pages} {33} (\bibinfo {year} {2017})}\BibitemShut {NoStop}%
\bibitem [{\citenamefont {Ohno}\ \emph {et~al.}(2000)\citenamefont {Ohno},
  \citenamefont {Chiba}, \citenamefont {Matsukura}, \citenamefont {Omiya},
  \citenamefont {Abe}, \citenamefont {Dietl}, \citenamefont {Ohno},\ and\
  \citenamefont {Ohtani}}]{ohno_2000}%
  \BibitemOpen
  \bibfield  {author} {\bibinfo {author} {\bibfnamefont {H.}~\bibnamefont
  {Ohno}}, \bibinfo {author} {\bibfnamefont {D.}~\bibnamefont {Chiba}},
  \bibinfo {author} {\bibfnamefont {F.}~\bibnamefont {Matsukura}}, \bibinfo
  {author} {\bibfnamefont {T.}~\bibnamefont {Omiya}}, \bibinfo {author}
  {\bibfnamefont {E.}~\bibnamefont {Abe}}, \bibinfo {author} {\bibfnamefont
  {T.}~\bibnamefont {Dietl}}, \bibinfo {author} {\bibfnamefont
  {Y.}~\bibnamefont {Ohno}}, \ and\ \bibinfo {author} {\bibfnamefont
  {K.}~\bibnamefont {Ohtani}},\ }\href {\doibase 10.1038/35050040} {\bibfield
  {journal} {\bibinfo  {journal} {Nature}\ }\textbf {\bibinfo {volume} {408}},\
  \bibinfo {pages} {944} (\bibinfo {year} {2000})}\BibitemShut {NoStop}%
\bibitem [{\citenamefont {Chiba}\ \emph {et~al.}(2011)\citenamefont {Chiba},
  \citenamefont {Fukami}, \citenamefont {Shimamura}, \citenamefont {Ishiwata},
  \citenamefont {Kobayashi},\ and\ \citenamefont {Ono}}]{chiba_2011}%
  \BibitemOpen
  \bibfield  {author} {\bibinfo {author} {\bibfnamefont {D.}~\bibnamefont
  {Chiba}}, \bibinfo {author} {\bibfnamefont {S.}~\bibnamefont {Fukami}},
  \bibinfo {author} {\bibfnamefont {K.}~\bibnamefont {Shimamura}}, \bibinfo
  {author} {\bibfnamefont {N.}~\bibnamefont {Ishiwata}}, \bibinfo {author}
  {\bibfnamefont {K.}~\bibnamefont {Kobayashi}}, \ and\ \bibinfo {author}
  {\bibfnamefont {T.}~\bibnamefont {Ono}},\ }\href {\doibase 10.1038/nmat3130}
  {\bibfield  {journal} {\bibinfo  {journal} {Nature Materials}\ }\textbf
  {\bibinfo {volume} {10}},\ \bibinfo {pages} {853} (\bibinfo {year}
  {2011})}\BibitemShut {NoStop}%
\bibitem [{\citenamefont {Chiba}\ \emph {et~al.}(2003)\citenamefont {Chiba},
  \citenamefont {Yamanouchi}, \citenamefont {Matsukura},\ and\ \citenamefont
  {Ohno}}]{chiba_2003}%
  \BibitemOpen
  \bibfield  {author} {\bibinfo {author} {\bibfnamefont {D.}~\bibnamefont
  {Chiba}}, \bibinfo {author} {\bibfnamefont {M.}~\bibnamefont {Yamanouchi}},
  \bibinfo {author} {\bibfnamefont {F.}~\bibnamefont {Matsukura}}, \ and\
  \bibinfo {author} {\bibfnamefont {H.}~\bibnamefont {Ohno}},\ }\href {\doibase
  10.1126/science.1086608} {\bibfield  {journal} {\bibinfo  {journal}
  {Science}\ }\textbf {\bibinfo {volume} {301}},\ \bibinfo {pages} {943}
  (\bibinfo {year} {2003})}\BibitemShut {NoStop}%
\bibitem [{\citenamefont {Weisheit}\ \emph {et~al.}(2007)\citenamefont
  {Weisheit}, \citenamefont {F{\"a}hler}, \citenamefont {Marty}, \citenamefont
  {Souche}, \citenamefont {Poinsignon},\ and\ \citenamefont
  {Givord}}]{weisheit_2007}%
  \BibitemOpen
  \bibfield  {author} {\bibinfo {author} {\bibfnamefont {M.}~\bibnamefont
  {Weisheit}}, \bibinfo {author} {\bibfnamefont {S.}~\bibnamefont
  {F{\"a}hler}}, \bibinfo {author} {\bibfnamefont {A.}~\bibnamefont {Marty}},
  \bibinfo {author} {\bibfnamefont {Y.}~\bibnamefont {Souche}}, \bibinfo
  {author} {\bibfnamefont {C.}~\bibnamefont {Poinsignon}}, \ and\ \bibinfo
  {author} {\bibfnamefont {D.}~\bibnamefont {Givord}},\ }\href {\doibase
  10.1126/science.1136629} {\bibfield  {journal} {\bibinfo  {journal}
  {Science}\ }\textbf {\bibinfo {volume} {315}},\ \bibinfo {pages} {349}
  (\bibinfo {year} {2007})}\BibitemShut {NoStop}%
\bibitem [{\citenamefont {Endo}\ \emph {et~al.}(2010)\citenamefont {Endo},
  \citenamefont {Kanai}, \citenamefont {Ikeda}, \citenamefont {Matsukura},\
  and\ \citenamefont {Ohno}}]{endo_2010}%
  \BibitemOpen
  \bibfield  {author} {\bibinfo {author} {\bibfnamefont {M.}~\bibnamefont
  {Endo}}, \bibinfo {author} {\bibfnamefont {S.}~\bibnamefont {Kanai}},
  \bibinfo {author} {\bibfnamefont {S.}~\bibnamefont {Ikeda}}, \bibinfo
  {author} {\bibfnamefont {F.}~\bibnamefont {Matsukura}}, \ and\ \bibinfo
  {author} {\bibfnamefont {H.}~\bibnamefont {Ohno}},\ }\href {\doibase
  doi:10.1063/1.3429592} {\bibfield  {journal} {\bibinfo  {journal} {Applied
  Physics Letters}\ }\textbf {\bibinfo {volume} {96}},\ \bibinfo {pages}
  {212503} (\bibinfo {year} {2010})}\BibitemShut {NoStop}%
\bibitem [{\citenamefont {Obinata}\ \emph {et~al.}(2015)\citenamefont
  {Obinata}, \citenamefont {Hibino}, \citenamefont {Hayakawa}, \citenamefont
  {Koyama}, \citenamefont {Miwa}, \citenamefont {Ono},\ and\ \citenamefont
  {Chiba}}]{obinata_2015}%
  \BibitemOpen
  \bibfield  {author} {\bibinfo {author} {\bibfnamefont {A.}~\bibnamefont
  {Obinata}}, \bibinfo {author} {\bibfnamefont {Y.}~\bibnamefont {Hibino}},
  \bibinfo {author} {\bibfnamefont {D.}~\bibnamefont {Hayakawa}}, \bibinfo
  {author} {\bibfnamefont {T.}~\bibnamefont {Koyama}}, \bibinfo {author}
  {\bibfnamefont {K.}~\bibnamefont {Miwa}}, \bibinfo {author} {\bibfnamefont
  {S.}~\bibnamefont {Ono}}, \ and\ \bibinfo {author} {\bibfnamefont
  {D.}~\bibnamefont {Chiba}},\ }\href {\doibase 10.1038/srep14303} {\bibfield
  {journal} {\bibinfo  {journal} {Scientific Reports}\ }\textbf {\bibinfo
  {volume} {5}} (\bibinfo {year} {2015}),\ 10.1038/srep14303}\BibitemShut
  {NoStop}%
\bibitem [{\citenamefont {Garcia}\ \emph {et~al.}(2010)\citenamefont {Garcia},
  \citenamefont {Bibes}, \citenamefont {Bocher}, \citenamefont {Valencia},
  \citenamefont {Kronast}, \citenamefont {Crassous}, \citenamefont {Moya},
  \citenamefont {Enouz-Vedrenne}, \citenamefont {Gloter}, \citenamefont
  {Imhoff}, \citenamefont {Deranlot}, \citenamefont {Mathur}, \citenamefont
  {Fusil}, \citenamefont {Bouzehouane},\ and\ \citenamefont
  {Barth{\'e}l{\'e}my}}]{garcia_2010}%
  \BibitemOpen
  \bibfield  {author} {\bibinfo {author} {\bibfnamefont {V.}~\bibnamefont
  {Garcia}}, \bibinfo {author} {\bibfnamefont {M.}~\bibnamefont {Bibes}},
  \bibinfo {author} {\bibfnamefont {L.}~\bibnamefont {Bocher}}, \bibinfo
  {author} {\bibfnamefont {S.}~\bibnamefont {Valencia}}, \bibinfo {author}
  {\bibfnamefont {F.}~\bibnamefont {Kronast}}, \bibinfo {author} {\bibfnamefont
  {A.}~\bibnamefont {Crassous}}, \bibinfo {author} {\bibfnamefont
  {X.}~\bibnamefont {Moya}}, \bibinfo {author} {\bibfnamefont {S.}~\bibnamefont
  {Enouz-Vedrenne}}, \bibinfo {author} {\bibfnamefont {A.}~\bibnamefont
  {Gloter}}, \bibinfo {author} {\bibfnamefont {D.}~\bibnamefont {Imhoff}},
  \bibinfo {author} {\bibfnamefont {C.}~\bibnamefont {Deranlot}}, \bibinfo
  {author} {\bibfnamefont {N.~D.}\ \bibnamefont {Mathur}}, \bibinfo {author}
  {\bibfnamefont {S.}~\bibnamefont {Fusil}}, \bibinfo {author} {\bibfnamefont
  {K.}~\bibnamefont {Bouzehouane}}, \ and\ \bibinfo {author} {\bibfnamefont
  {A.}~\bibnamefont {Barth{\'e}l{\'e}my}},\ }\href {\doibase
  10.1126/science.1184028} {\bibfield  {journal} {\bibinfo  {journal}
  {Science}\ }\textbf {\bibinfo {volume} {327}},\ \bibinfo {pages} {1106}
  (\bibinfo {year} {2010})}\BibitemShut {NoStop}%
\bibitem [{\citenamefont {Chiba}\ \emph {et~al.}(2008)\citenamefont {Chiba},
  \citenamefont {Sawicki}, \citenamefont {Nishitani}, \citenamefont {Nakatani},
  \citenamefont {Matsukura},\ and\ \citenamefont {Ohno}}]{chiba_2008}%
  \BibitemOpen
  \bibfield  {author} {\bibinfo {author} {\bibfnamefont {D.}~\bibnamefont
  {Chiba}}, \bibinfo {author} {\bibfnamefont {M.}~\bibnamefont {Sawicki}},
  \bibinfo {author} {\bibfnamefont {Y.}~\bibnamefont {Nishitani}}, \bibinfo
  {author} {\bibfnamefont {Y.}~\bibnamefont {Nakatani}}, \bibinfo {author}
  {\bibfnamefont {F.}~\bibnamefont {Matsukura}}, \ and\ \bibinfo {author}
  {\bibfnamefont {H.}~\bibnamefont {Ohno}},\ }\href {\doibase
  10.1038/nature07318} {\bibfield  {journal} {\bibinfo  {journal} {Nature}\
  }\textbf {\bibinfo {volume} {455}},\ \bibinfo {pages} {515} (\bibinfo {year}
  {2008})}\BibitemShut {NoStop}%
\bibitem [{\citenamefont {Maruyama}\ \emph
  {et~al.}(2009{\natexlab{a}})\citenamefont {Maruyama}, \citenamefont {Shiota},
  \citenamefont {Nozaki}, \citenamefont {Ohta}, \citenamefont {Toda},
  \citenamefont {Mizuguchi}, \citenamefont {Tulapurkar}, \citenamefont
  {Shinjo}, \citenamefont {Shiraishi}, \citenamefont {Mizukami}, \citenamefont
  {Ando},\ and\ \citenamefont {Suzuki}}]{maruyama_2009}%
  \BibitemOpen
  \bibfield  {author} {\bibinfo {author} {\bibfnamefont {T.}~\bibnamefont
  {Maruyama}}, \bibinfo {author} {\bibfnamefont {Y.}~\bibnamefont {Shiota}},
  \bibinfo {author} {\bibfnamefont {T.}~\bibnamefont {Nozaki}}, \bibinfo
  {author} {\bibfnamefont {K.}~\bibnamefont {Ohta}}, \bibinfo {author}
  {\bibfnamefont {N.}~\bibnamefont {Toda}}, \bibinfo {author} {\bibfnamefont
  {M.}~\bibnamefont {Mizuguchi}}, \bibinfo {author} {\bibfnamefont {A.~A.}\
  \bibnamefont {Tulapurkar}}, \bibinfo {author} {\bibfnamefont
  {T.}~\bibnamefont {Shinjo}}, \bibinfo {author} {\bibfnamefont
  {M.}~\bibnamefont {Shiraishi}}, \bibinfo {author} {\bibfnamefont
  {S.}~\bibnamefont {Mizukami}}, \bibinfo {author} {\bibfnamefont
  {Y.}~\bibnamefont {Ando}}, \ and\ \bibinfo {author} {\bibfnamefont
  {Y.}~\bibnamefont {Suzuki}},\ }\href {\doibase 10.1038/nnano.2008.406}
  {\bibfield  {journal} {\bibinfo  {journal} {Nature Nanotechnology}\ }\textbf
  {\bibinfo {volume} {4}},\ \bibinfo {pages} {158} (\bibinfo {year}
  {2009}{\natexlab{a}})}\BibitemShut {NoStop}%
\bibitem [{\citenamefont {Shiota}\ \emph {et~al.}(2011)\citenamefont {Shiota},
  \citenamefont {Nozaki}, \citenamefont {Bonell}, \citenamefont {Murakami},
  \citenamefont {Shinjo},\ and\ \citenamefont {Suzuki}}]{shiota_2011}%
  \BibitemOpen
  \bibfield  {author} {\bibinfo {author} {\bibfnamefont {Y.}~\bibnamefont
  {Shiota}}, \bibinfo {author} {\bibfnamefont {T.}~\bibnamefont {Nozaki}},
  \bibinfo {author} {\bibfnamefont {F.}~\bibnamefont {Bonell}}, \bibinfo
  {author} {\bibfnamefont {S.}~\bibnamefont {Murakami}}, \bibinfo {author}
  {\bibfnamefont {T.}~\bibnamefont {Shinjo}}, \ and\ \bibinfo {author}
  {\bibfnamefont {Y.}~\bibnamefont {Suzuki}},\ }\href {\doibase
  10.1038/nmat3172} {\bibfield  {journal} {\bibinfo  {journal} {Nature
  Materials}\ }\textbf {\bibinfo {volume} {11}},\ \bibinfo {pages} {39}
  (\bibinfo {year} {2011})}\BibitemShut {NoStop}%
\bibitem [{\citenamefont {Wang}\ \emph {et~al.}(2012)\citenamefont {Wang},
  \citenamefont {Li}, \citenamefont {Hageman},\ and\ \citenamefont
  {Chien}}]{wang_2012}%
  \BibitemOpen
  \bibfield  {author} {\bibinfo {author} {\bibfnamefont {W.-G.}\ \bibnamefont
  {Wang}}, \bibinfo {author} {\bibfnamefont {M.}~\bibnamefont {Li}}, \bibinfo
  {author} {\bibfnamefont {S.}~\bibnamefont {Hageman}}, \ and\ \bibinfo
  {author} {\bibfnamefont {C.~L.}\ \bibnamefont {Chien}},\ }\href {\doibase
  10.1038/nmat3171} {\bibfield  {journal} {\bibinfo  {journal} {Nature
  Materials}\ }\textbf {\bibinfo {volume} {11}},\ \bibinfo {pages} {64}
  (\bibinfo {year} {2012})}\BibitemShut {NoStop}%
\bibitem [{\citenamefont {Kanai}\ \emph {et~al.}(2012)\citenamefont {Kanai},
  \citenamefont {Yamanouchi}, \citenamefont {Ikeda}, \citenamefont {Nakatani},
  \citenamefont {Matsukura},\ and\ \citenamefont {Ohno}}]{kanai_2012}%
  \BibitemOpen
  \bibfield  {author} {\bibinfo {author} {\bibfnamefont {S.}~\bibnamefont
  {Kanai}}, \bibinfo {author} {\bibfnamefont {M.}~\bibnamefont {Yamanouchi}},
  \bibinfo {author} {\bibfnamefont {S.}~\bibnamefont {Ikeda}}, \bibinfo
  {author} {\bibfnamefont {Y.}~\bibnamefont {Nakatani}}, \bibinfo {author}
  {\bibfnamefont {F.}~\bibnamefont {Matsukura}}, \ and\ \bibinfo {author}
  {\bibfnamefont {H.}~\bibnamefont {Ohno}},\ }\href {\doibase
  10.1063/1.4753816} {\bibfield  {journal} {\bibinfo  {journal} {Applied
  Physics Letters}\ }\textbf {\bibinfo {volume} {101}},\ \bibinfo {pages}
  {122403} (\bibinfo {year} {2012})}\BibitemShut {NoStop}%
\bibitem [{\citenamefont {Yang}\ \emph {et~al.}(2011)\citenamefont {Yang},
  \citenamefont {Chshiev}, \citenamefont {Dieny}, \citenamefont {Lee},
  \citenamefont {Manchon},\ and\ \citenamefont {Shin}}]{yang2011}%
  \BibitemOpen
  \bibfield  {author} {\bibinfo {author} {\bibfnamefont {H.~X.}\ \bibnamefont
  {Yang}}, \bibinfo {author} {\bibfnamefont {M.}~\bibnamefont {Chshiev}},
  \bibinfo {author} {\bibfnamefont {B.}~\bibnamefont {Dieny}}, \bibinfo
  {author} {\bibfnamefont {J.~H.}\ \bibnamefont {Lee}}, \bibinfo {author}
  {\bibfnamefont {A.}~\bibnamefont {Manchon}}, \ and\ \bibinfo {author}
  {\bibfnamefont {K.~H.}\ \bibnamefont {Shin}},\ }\href {\doibase
  10.1103/PhysRevB.84.054401} {\bibfield  {journal} {\bibinfo  {journal}
  {Physical Review B}\ }\textbf {\bibinfo {volume} {84}},\ \bibinfo {pages}
  {054401} (\bibinfo {year} {2011})}\BibitemShut {NoStop}%
\bibitem [{\citenamefont {Amiri}\ \emph {et~al.}(2015)\citenamefont {Amiri},
  \citenamefont {Alzate}, \citenamefont {Cai}, \citenamefont {Ebrahimi},
  \citenamefont {Hu}, \citenamefont {Wong}, \citenamefont {Gr\`ezes},
  \citenamefont {Lee}, \citenamefont {Yu}, \citenamefont {Li}, \citenamefont
  {Akyol}, \citenamefont {Shao}, \citenamefont {Katine}, \citenamefont
  {Langer}, \citenamefont {Ocker},\ and\ \citenamefont {Wang}}]{amiri_2015}%
  \BibitemOpen
  \bibfield  {author} {\bibinfo {author} {\bibfnamefont {P.~K.}\ \bibnamefont
  {Amiri}}, \bibinfo {author} {\bibfnamefont {J.~G.}\ \bibnamefont {Alzate}},
  \bibinfo {author} {\bibfnamefont {X.~Q.}\ \bibnamefont {Cai}}, \bibinfo
  {author} {\bibfnamefont {F.}~\bibnamefont {Ebrahimi}}, \bibinfo {author}
  {\bibfnamefont {Q.}~\bibnamefont {Hu}}, \bibinfo {author} {\bibfnamefont
  {K.}~\bibnamefont {Wong}}, \bibinfo {author} {\bibfnamefont {C.}~\bibnamefont
  {Gr\`ezes}}, \bibinfo {author} {\bibfnamefont {H.}~\bibnamefont {Lee}},
  \bibinfo {author} {\bibfnamefont {G.}~\bibnamefont {Yu}}, \bibinfo {author}
  {\bibfnamefont {X.}~\bibnamefont {Li}}, \bibinfo {author} {\bibfnamefont
  {M.}~\bibnamefont {Akyol}}, \bibinfo {author} {\bibfnamefont
  {Q.}~\bibnamefont {Shao}}, \bibinfo {author} {\bibfnamefont {J.~A.}\
  \bibnamefont {Katine}}, \bibinfo {author} {\bibfnamefont {J.}~\bibnamefont
  {Langer}}, \bibinfo {author} {\bibfnamefont {B.}~\bibnamefont {Ocker}}, \
  and\ \bibinfo {author} {\bibfnamefont {K.~L.}\ \bibnamefont {Wang}},\ }\href
  {\doibase 10.1109/TMAG.2015.2443124} {\bibfield  {journal} {\bibinfo
  {journal} {IEEE Transactions on Magnetics}\ }\textbf {\bibinfo {volume}
  {51}},\ \bibinfo {pages} {1} (\bibinfo {year} {2015})}\BibitemShut {NoStop}%
\bibitem [{\citenamefont {Grezes}\ \emph {et~al.}(2017)\citenamefont {Grezes},
  \citenamefont {Lee}, \citenamefont {Lee}, \citenamefont {Wang}, \citenamefont
  {Ebrahimi}, \citenamefont {Li}, \citenamefont {Wong}, \citenamefont {Katine},
  \citenamefont {Ocker}, \citenamefont {Langer}, \citenamefont {Gupta},
  \citenamefont {Amiri},\ and\ \citenamefont {Wang}}]{grezes_2017}%
  \BibitemOpen
  \bibfield  {author} {\bibinfo {author} {\bibfnamefont {C.}~\bibnamefont
  {Grezes}}, \bibinfo {author} {\bibfnamefont {H.}~\bibnamefont {Lee}},
  \bibinfo {author} {\bibfnamefont {A.}~\bibnamefont {Lee}}, \bibinfo {author}
  {\bibfnamefont {S.}~\bibnamefont {Wang}}, \bibinfo {author} {\bibfnamefont
  {F.}~\bibnamefont {Ebrahimi}}, \bibinfo {author} {\bibfnamefont
  {X.}~\bibnamefont {Li}}, \bibinfo {author} {\bibfnamefont {K.}~\bibnamefont
  {Wong}}, \bibinfo {author} {\bibfnamefont {J.~A.}\ \bibnamefont {Katine}},
  \bibinfo {author} {\bibfnamefont {B.}~\bibnamefont {Ocker}}, \bibinfo
  {author} {\bibfnamefont {J.}~\bibnamefont {Langer}}, \bibinfo {author}
  {\bibfnamefont {P.}~\bibnamefont {Gupta}}, \bibinfo {author} {\bibfnamefont
  {P.~K.}\ \bibnamefont {Amiri}}, \ and\ \bibinfo {author} {\bibfnamefont
  {K.~L.}\ \bibnamefont {Wang}},\ }\href {\doibase 10.1109/LMAG.2016.2630667}
  {\bibfield  {journal} {\bibinfo  {journal} {IEEE Magnetics Letters}\ }\textbf
  {\bibinfo {volume} {8}},\ \bibinfo {pages} {1} (\bibinfo {year}
  {2017})}\BibitemShut {NoStop}%
\bibitem [{\citenamefont {Shiota}\ \emph {et~al.}(2017)\citenamefont {Shiota},
  \citenamefont {Nozaki}, \citenamefont {Tamaru}, \citenamefont {Yakushiji},
  \citenamefont {Kubota}, \citenamefont {Fukushima}, \citenamefont {Yuasa},\
  and\ \citenamefont {Suzuki}}]{shiota_2017}%
  \BibitemOpen
  \bibfield  {author} {\bibinfo {author} {\bibfnamefont {Y.}~\bibnamefont
  {Shiota}}, \bibinfo {author} {\bibfnamefont {T.}~\bibnamefont {Nozaki}},
  \bibinfo {author} {\bibfnamefont {S.}~\bibnamefont {Tamaru}}, \bibinfo
  {author} {\bibfnamefont {K.}~\bibnamefont {Yakushiji}}, \bibinfo {author}
  {\bibfnamefont {H.}~\bibnamefont {Kubota}}, \bibinfo {author} {\bibfnamefont
  {A.}~\bibnamefont {Fukushima}}, \bibinfo {author} {\bibfnamefont
  {S.}~\bibnamefont {Yuasa}}, \ and\ \bibinfo {author} {\bibfnamefont
  {Y.}~\bibnamefont {Suzuki}},\ }\href {\doibase 10.1063/1.4990680} {\bibfield
  {journal} {\bibinfo  {journal} {Applied Physics Letters}\ }\textbf {\bibinfo
  {volume} {111}},\ \bibinfo {pages} {022408} (\bibinfo {year}
  {2017})}\BibitemShut {NoStop}%
\bibitem [{\citenamefont {Duan}\ \emph {et~al.}(2008)\citenamefont {Duan},
  \citenamefont {Velev}, \citenamefont {Sabirianov}, \citenamefont {Mei},
  \citenamefont {Jaswal},\ and\ \citenamefont {Tsymbal}}]{duan_2008a}%
  \BibitemOpen
  \bibfield  {author} {\bibinfo {author} {\bibfnamefont {C.-G.}\ \bibnamefont
  {Duan}}, \bibinfo {author} {\bibfnamefont {J.~P.}\ \bibnamefont {Velev}},
  \bibinfo {author} {\bibfnamefont {R.~F.}\ \bibnamefont {Sabirianov}},
  \bibinfo {author} {\bibfnamefont {W.~N.}\ \bibnamefont {Mei}}, \bibinfo
  {author} {\bibfnamefont {S.~S.}\ \bibnamefont {Jaswal}}, \ and\ \bibinfo
  {author} {\bibfnamefont {E.~Y.}\ \bibnamefont {Tsymbal}},\ }\href {\doibase
  10.1063/1.2901879} {\bibfield  {journal} {\bibinfo  {journal} {Applied
  Physics Letters}\ }\textbf {\bibinfo {volume} {92}},\ \bibinfo {pages}
  {122905} (\bibinfo {year} {2008})}\BibitemShut {NoStop}%
\bibitem [{\citenamefont {Mardana}\ \emph {et~al.}(2011)\citenamefont
  {Mardana}, \citenamefont {Ducharme},\ and\ \citenamefont
  {Adenwalla}}]{mardana_2011}%
  \BibitemOpen
  \bibfield  {author} {\bibinfo {author} {\bibfnamefont {A.}~\bibnamefont
  {Mardana}}, \bibinfo {author} {\bibfnamefont {S.}~\bibnamefont {Ducharme}}, \
  and\ \bibinfo {author} {\bibfnamefont {S.}~\bibnamefont {Adenwalla}},\ }\href
  {\doibase 10.1021/nl201965r} {\bibfield  {journal} {\bibinfo  {journal} {Nano
  Letters}\ }\textbf {\bibinfo {volume} {11}},\ \bibinfo {pages} {3862}
  (\bibinfo {year} {2011})}\BibitemShut {NoStop}%
\bibitem [{\citenamefont {Lukashev}\ \emph
  {et~al.}(2012{\natexlab{a}})\citenamefont {Lukashev}, \citenamefont {Burton},
  \citenamefont {Jaswal},\ and\ \citenamefont {Tsymbal}}]{lukashev2012}%
  \BibitemOpen
  \bibfield  {author} {\bibinfo {author} {\bibfnamefont {P.~V.}\ \bibnamefont
  {Lukashev}}, \bibinfo {author} {\bibfnamefont {J.~D.}\ \bibnamefont
  {Burton}}, \bibinfo {author} {\bibfnamefont {S.~S.}\ \bibnamefont {Jaswal}},
  \ and\ \bibinfo {author} {\bibfnamefont {E.~Y.}\ \bibnamefont {Tsymbal}},\
  }\href {\doibase 10.1088/0953-8984/24/22/226003} {\bibfield  {journal}
  {\bibinfo  {journal} {Journal of Physics: Condensed Matter}\ }\textbf
  {\bibinfo {volume} {24}},\ \bibinfo {pages} {226003} (\bibinfo {year}
  {2012}{\natexlab{a}})}\BibitemShut {NoStop}%
\bibitem [{\citenamefont {Lukashev}\ \emph
  {et~al.}(2012{\natexlab{b}})\citenamefont {Lukashev}, \citenamefont {Paudel},
  \citenamefont {{L\'opez-Encarnaci\'on}}, \citenamefont {Adenwalla},
  \citenamefont {Tsymbal},\ and\ \citenamefont {Velev}}]{lukashev_2012a}%
  \BibitemOpen
  \bibfield  {author} {\bibinfo {author} {\bibfnamefont {P.~V.}\ \bibnamefont
  {Lukashev}}, \bibinfo {author} {\bibfnamefont {T.~R.}\ \bibnamefont
  {Paudel}}, \bibinfo {author} {\bibfnamefont {J.~M.}\ \bibnamefont
  {{L\'opez-Encarnaci\'on}}}, \bibinfo {author} {\bibfnamefont
  {S.}~\bibnamefont {Adenwalla}}, \bibinfo {author} {\bibfnamefont {E.~Y.}\
  \bibnamefont {Tsymbal}}, \ and\ \bibinfo {author} {\bibfnamefont {J.~P.}\
  \bibnamefont {Velev}},\ }\href {\doibase 10.1021/nn303212h} {\bibfield
  {journal} {\bibinfo  {journal} {ACS Nano}\ }\textbf {\bibinfo {volume} {6}},\
  \bibinfo {pages} {9745} (\bibinfo {year} {2012}{\natexlab{b}})}\BibitemShut
  {NoStop}%
\bibitem [{\citenamefont {Lee}\ \emph {et~al.}(2013)\citenamefont {Lee},
  \citenamefont {Choi},\ and\ \citenamefont {Chung}}]{lee_2013b}%
  \BibitemOpen
  \bibfield  {author} {\bibinfo {author} {\bibfnamefont {M.}~\bibnamefont
  {Lee}}, \bibinfo {author} {\bibfnamefont {H.}~\bibnamefont {Choi}}, \ and\
  \bibinfo {author} {\bibfnamefont {Y.-C.}\ \bibnamefont {Chung}},\ }\href
  {\doibase 10.1063/1.4800499} {\bibfield  {journal} {\bibinfo  {journal}
  {Journal of Applied Physics}\ }\textbf {\bibinfo {volume} {113}},\ \bibinfo
  {pages} {17C729} (\bibinfo {year} {2013})}\BibitemShut {NoStop}%
\bibitem [{\citenamefont {Grange}\ \emph {et~al.}(1998)\citenamefont {Grange},
  \citenamefont {Maret}, \citenamefont {Kappler}, \citenamefont {Vogel},
  \citenamefont {Fontaine}, \citenamefont {Petroff}, \citenamefont {Krill},
  \citenamefont {Rogalev}, \citenamefont {Goulon}, \citenamefont {Finazzi},\
  and\ \citenamefont {Brookes}}]{grange_1998a}%
  \BibitemOpen
  \bibfield  {author} {\bibinfo {author} {\bibfnamefont {W.}~\bibnamefont
  {Grange}}, \bibinfo {author} {\bibfnamefont {M.}~\bibnamefont {Maret}},
  \bibinfo {author} {\bibfnamefont {J.-P.}\ \bibnamefont {Kappler}}, \bibinfo
  {author} {\bibfnamefont {J.}~\bibnamefont {Vogel}}, \bibinfo {author}
  {\bibfnamefont {A.}~\bibnamefont {Fontaine}}, \bibinfo {author}
  {\bibfnamefont {F.}~\bibnamefont {Petroff}}, \bibinfo {author} {\bibfnamefont
  {G.}~\bibnamefont {Krill}}, \bibinfo {author} {\bibfnamefont
  {A.}~\bibnamefont {Rogalev}}, \bibinfo {author} {\bibfnamefont
  {J.}~\bibnamefont {Goulon}}, \bibinfo {author} {\bibfnamefont
  {M.}~\bibnamefont {Finazzi}}, \ and\ \bibinfo {author} {\bibfnamefont
  {N.~B.}\ \bibnamefont {Brookes}},\ }\href {\doibase 10.1103/PhysRevB.58.6298}
  {\bibfield  {journal} {\bibinfo  {journal} {Physical Review B}\ }\textbf
  {\bibinfo {volume} {58}},\ \bibinfo {pages} {6298} (\bibinfo {year}
  {1998})}\BibitemShut {NoStop}%
\bibitem [{\citenamefont {Mryasov}\ \emph {et~al.}(2005)\citenamefont
  {Mryasov}, \citenamefont {Nowak}, \citenamefont {Guslienko},\ and\
  \citenamefont {Chantrell}}]{mryasov_2005}%
  \BibitemOpen
  \bibfield  {author} {\bibinfo {author} {\bibfnamefont {O.~N.}\ \bibnamefont
  {Mryasov}}, \bibinfo {author} {\bibfnamefont {U.}~\bibnamefont {Nowak}},
  \bibinfo {author} {\bibfnamefont {K.~Y.}\ \bibnamefont {Guslienko}}, \ and\
  \bibinfo {author} {\bibfnamefont {R.~W.}\ \bibnamefont {Chantrell}},\ }\href
  {\doibase 10.1209/epl/i2004-10404-2} {\bibfield  {journal} {\bibinfo
  {journal} {EPL (Europhysics Letters)}\ }\textbf {\bibinfo {volume} {69}},\
  \bibinfo {pages} {805} (\bibinfo {year} {2005})}\BibitemShut {NoStop}%
\bibitem [{\citenamefont {Shick}\ \emph {et~al.}(2008)\citenamefont {Shick},
  \citenamefont {M{\'a}ca}, \citenamefont {Ondr{\'a}{\v c}ek}, \citenamefont
  {Mryasov},\ and\ \citenamefont {Jungwirth}}]{shick_2008}%
  \BibitemOpen
  \bibfield  {author} {\bibinfo {author} {\bibfnamefont {A.~B.}\ \bibnamefont
  {Shick}}, \bibinfo {author} {\bibfnamefont {F.}~\bibnamefont {M{\'a}ca}},
  \bibinfo {author} {\bibfnamefont {M.}~\bibnamefont {Ondr{\'a}{\v c}ek}},
  \bibinfo {author} {\bibfnamefont {O.~N.}\ \bibnamefont {Mryasov}}, \ and\
  \bibinfo {author} {\bibfnamefont {T.}~\bibnamefont {Jungwirth}},\ }\href
  {\doibase 10.1103/PhysRevB.78.054413} {\bibfield  {journal} {\bibinfo
  {journal} {Physical Review B}\ }\textbf {\bibinfo {volume} {78}},\ \bibinfo
  {pages} {054413} (\bibinfo {year} {2008})}\BibitemShut {NoStop}%
\bibitem [{\citenamefont {Sakuma}(1994)}]{sakuma_1994}%
  \BibitemOpen
  \bibfield  {author} {\bibinfo {author} {\bibfnamefont {A.}~\bibnamefont
  {Sakuma}},\ }\href {\doibase 10.1143/JPSJ.63.3053} {\bibfield  {journal}
  {\bibinfo  {journal} {Journal of the Physical Society of Japan}\ }\textbf
  {\bibinfo {volume} {63}},\ \bibinfo {pages} {3053} (\bibinfo {year}
  {1994})}\BibitemShut {NoStop}%
\bibitem [{\citenamefont {Weller}\ \emph {et~al.}(1992)\citenamefont {Weller},
  \citenamefont {Br{\"a}ndle}, \citenamefont {Gorman}, \citenamefont {Lin},\
  and\ \citenamefont {Notarys}}]{weller_1992}%
  \BibitemOpen
  \bibfield  {author} {\bibinfo {author} {\bibfnamefont {D.}~\bibnamefont
  {Weller}}, \bibinfo {author} {\bibfnamefont {H.}~\bibnamefont {Br{\"a}ndle}},
  \bibinfo {author} {\bibfnamefont {G.}~\bibnamefont {Gorman}}, \bibinfo
  {author} {\bibfnamefont {C.-J.}\ \bibnamefont {Lin}}, \ and\ \bibinfo
  {author} {\bibfnamefont {H.}~\bibnamefont {Notarys}},\ }\href {\doibase
  doi:10.1063/1.108074} {\bibfield  {journal} {\bibinfo  {journal} {Applied
  Physics Letters}\ }\textbf {\bibinfo {volume} {61}},\ \bibinfo {pages} {2726}
  (\bibinfo {year} {1992})}\BibitemShut {NoStop}%
\bibitem [{\citenamefont {Kozuka}\ \emph {et~al.}(2014)\citenamefont {Kozuka},
  \citenamefont {Tsukazaki},\ and\ \citenamefont {Kawasaki}}]{kozuka2014}%
  \BibitemOpen
  \bibfield  {author} {\bibinfo {author} {\bibfnamefont {Y.}~\bibnamefont
  {Kozuka}}, \bibinfo {author} {\bibfnamefont {A.}~\bibnamefont {Tsukazaki}}, \
  and\ \bibinfo {author} {\bibfnamefont {M.}~\bibnamefont {Kawasaki}},\ }\href
  {\doibase 10.1063/1.4853535} {\bibfield  {journal} {\bibinfo  {journal}
  {Applied Physics Reviews}\ }\textbf {\bibinfo {volume} {1}},\ \bibinfo
  {pages} {011303} (\bibinfo {year} {2014})}\BibitemShut {NoStop}%
\bibitem [{\citenamefont {Shick}\ and\ \citenamefont
  {Mryasov}(2003)}]{shick_2003}%
  \BibitemOpen
  \bibfield  {author} {\bibinfo {author} {\bibfnamefont {A.~B.}\ \bibnamefont
  {Shick}}\ and\ \bibinfo {author} {\bibfnamefont {O.~N.}\ \bibnamefont
  {Mryasov}},\ }\href {\doibase 10.1103/PhysRevB.67.172407} {\bibfield
  {journal} {\bibinfo  {journal} {Physical Review B}\ }\textbf {\bibinfo
  {volume} {67}},\ \bibinfo {pages} {172407} (\bibinfo {year}
  {2003})}\BibitemShut {NoStop}%
\bibitem [{\citenamefont {Belmoubarik}\ \emph {et~al.}(2015)\citenamefont
  {Belmoubarik}, \citenamefont {Al-Mahdawi}, \citenamefont {Sato},
  \citenamefont {Nozaki},\ and\ \citenamefont {Sahashi}}]{belmoubarik2015}%
  \BibitemOpen
  \bibfield  {author} {\bibinfo {author} {\bibfnamefont {M.}~\bibnamefont
  {Belmoubarik}}, \bibinfo {author} {\bibfnamefont {M.}~\bibnamefont
  {Al-Mahdawi}}, \bibinfo {author} {\bibfnamefont {H.}~\bibnamefont {Sato}},
  \bibinfo {author} {\bibfnamefont {T.}~\bibnamefont {Nozaki}}, \ and\ \bibinfo
  {author} {\bibfnamefont {M.}~\bibnamefont {Sahashi}},\ }\href {\doibase
  10.1063/1.4923041} {\bibfield  {journal} {\bibinfo  {journal} {Applied
  Physics Letters}\ }\textbf {\bibinfo {volume} {106}},\ \bibinfo {pages}
  {252403} (\bibinfo {year} {2015})}\BibitemShut {NoStop}%
\bibitem [{\citenamefont {Belmoubarik}\ \emph {et~al.}(2016)\citenamefont
  {Belmoubarik}, \citenamefont {Al-Mahdawi}, \citenamefont {Obata},
  \citenamefont {Yoshikawa}, \citenamefont {Sato}, \citenamefont {Nozaki},
  \citenamefont {Oda},\ and\ \citenamefont {Sahashi}}]{belmoubarik2016}%
  \BibitemOpen
  \bibfield  {author} {\bibinfo {author} {\bibfnamefont {M.}~\bibnamefont
  {Belmoubarik}}, \bibinfo {author} {\bibfnamefont {M.}~\bibnamefont
  {Al-Mahdawi}}, \bibinfo {author} {\bibfnamefont {M.}~\bibnamefont {Obata}},
  \bibinfo {author} {\bibfnamefont {D.}~\bibnamefont {Yoshikawa}}, \bibinfo
  {author} {\bibfnamefont {H.}~\bibnamefont {Sato}}, \bibinfo {author}
  {\bibfnamefont {T.}~\bibnamefont {Nozaki}}, \bibinfo {author} {\bibfnamefont
  {T.}~\bibnamefont {Oda}}, \ and\ \bibinfo {author} {\bibfnamefont
  {M.}~\bibnamefont {Sahashi}},\ }\href {\doibase 10.1063/1.4966180} {\bibfield
   {journal} {\bibinfo  {journal} {Applied Physics Letters}\ }\textbf {\bibinfo
  {volume} {109}},\ \bibinfo {pages} {173507} (\bibinfo {year}
  {2016})}\BibitemShut {NoStop}%
\bibitem [{\citenamefont {{Hohenberg}}\ and\ \citenamefont
  {{Kohn}}(1964)}]{hohenberg1964}%
  \BibitemOpen
  \bibfield  {author} {\bibinfo {author} {\bibfnamefont {P.}~\bibnamefont
  {{Hohenberg}}}\ and\ \bibinfo {author} {\bibfnamefont {W.}~\bibnamefont
  {{Kohn}}},\ }\href {\doibase 10.1103/PhysRev.136.B864} {\bibfield  {journal}
  {\bibinfo  {journal} {Physical Review}\ }\textbf {\bibinfo {volume} {136}},\
  \bibinfo {pages} {B864} (\bibinfo {year} {1964})}\BibitemShut {NoStop}%
\bibitem [{\citenamefont {{Vanderbilt}}(1990)}]{vanderbilt1990}%
  \BibitemOpen
  \bibfield  {author} {\bibinfo {author} {\bibfnamefont {D.}~\bibnamefont
  {{Vanderbilt}}},\ }\href {\doibase 10.1103/PhysRevB.41.7892} {\bibfield
  {journal} {\bibinfo  {journal} {Physical Review B}\ }\textbf {\bibinfo
  {volume} {41}},\ \bibinfo {pages} {7892} (\bibinfo {year}
  {1990})}\BibitemShut {NoStop}%
\bibitem [{\citenamefont {Tsujikawa}\ \emph {et~al.}(2008)\citenamefont
  {Tsujikawa}, \citenamefont {Hosokawa},\ and\ \citenamefont
  {Oda}}]{tsujikawa_2008}%
  \BibitemOpen
  \bibfield  {author} {\bibinfo {author} {\bibfnamefont {M.}~\bibnamefont
  {Tsujikawa}}, \bibinfo {author} {\bibfnamefont {A.}~\bibnamefont {Hosokawa}},
  \ and\ \bibinfo {author} {\bibfnamefont {T.}~\bibnamefont {Oda}},\ }\href
  {\doibase 10.1103/PhysRevB.77.054413} {\bibfield  {journal} {\bibinfo
  {journal} {Physical Review B}\ }\textbf {\bibinfo {volume} {77}},\ \bibinfo
  {pages} {054413} (\bibinfo {year} {2008})}\BibitemShut {NoStop}%
\bibitem [{\citenamefont {Perdew}\ \emph {et~al.}(1992)\citenamefont {Perdew},
  \citenamefont {Chevary}, \citenamefont {Vosko}, \citenamefont {Jackson},
  \citenamefont {Pederson}, \citenamefont {Singh},\ and\ \citenamefont
  {Fiolhais}}]{perdew1992}%
  \BibitemOpen
  \bibfield  {author} {\bibinfo {author} {\bibfnamefont {J.~P.}\ \bibnamefont
  {Perdew}}, \bibinfo {author} {\bibfnamefont {J.~A.}\ \bibnamefont {Chevary}},
  \bibinfo {author} {\bibfnamefont {S.~H.}\ \bibnamefont {Vosko}}, \bibinfo
  {author} {\bibfnamefont {K.~A.}\ \bibnamefont {Jackson}}, \bibinfo {author}
  {\bibfnamefont {M.~R.}\ \bibnamefont {Pederson}}, \bibinfo {author}
  {\bibfnamefont {D.~J.}\ \bibnamefont {Singh}}, \ and\ \bibinfo {author}
  {\bibfnamefont {C.}~\bibnamefont {Fiolhais}},\ }\href {\doibase
  10.1103/PhysRevB.46.6671} {\bibfield  {journal} {\bibinfo  {journal}
  {Physical Review B}\ }\textbf {\bibinfo {volume} {46}},\ \bibinfo {pages}
  {6671} (\bibinfo {year} {1992})}\BibitemShut {NoStop}%
\bibitem [{\citenamefont {Maruyama}\ \emph
  {et~al.}(2009{\natexlab{b}})\citenamefont {Maruyama}, \citenamefont {Shiota},
  \citenamefont {Nozaki}, \citenamefont {Ohta}, \citenamefont {Toda},
  \citenamefont {Mizuguchi}, \citenamefont {Tulapurkar}, \citenamefont
  {Shinjo}, \citenamefont {Shiraishi}, \citenamefont {Mizukami}, \citenamefont
  {Ando},\ and\ \citenamefont {Suzuki}}]{maruyama2009}%
  \BibitemOpen
  \bibfield  {author} {\bibinfo {author} {\bibfnamefont {T.}~\bibnamefont
  {Maruyama}}, \bibinfo {author} {\bibfnamefont {Y.}~\bibnamefont {Shiota}},
  \bibinfo {author} {\bibfnamefont {T.}~\bibnamefont {Nozaki}}, \bibinfo
  {author} {\bibfnamefont {K.}~\bibnamefont {Ohta}}, \bibinfo {author}
  {\bibfnamefont {N.}~\bibnamefont {Toda}}, \bibinfo {author} {\bibfnamefont
  {M.}~\bibnamefont {Mizuguchi}}, \bibinfo {author} {\bibfnamefont {A.~A.}\
  \bibnamefont {Tulapurkar}}, \bibinfo {author} {\bibfnamefont
  {T.}~\bibnamefont {Shinjo}}, \bibinfo {author} {\bibfnamefont
  {M.}~\bibnamefont {Shiraishi}}, \bibinfo {author} {\bibfnamefont
  {S.}~\bibnamefont {Mizukami}}, \bibinfo {author} {\bibfnamefont
  {Y.}~\bibnamefont {Ando}}, \ and\ \bibinfo {author} {\bibfnamefont
  {Y.}~\bibnamefont {Suzuki}},\ }\href {\doibase 10.1038/nnano.2008.406}
  {\bibfield  {journal} {\bibinfo  {journal} {Nature Nanotechnology}\ }\textbf
  {\bibinfo {volume} {4}},\ \bibinfo {pages} {158} (\bibinfo {year}
  {2009}{\natexlab{b}})}\BibitemShut {NoStop}%
\bibitem [{\citenamefont {Nakamura}\ \emph {et~al.}(2010)\citenamefont
  {Nakamura}, \citenamefont {Akiyama}, \citenamefont {Ito}, \citenamefont
  {Weinert},\ and\ \citenamefont {Freeman}}]{nakamura2010}%
  \BibitemOpen
  \bibfield  {author} {\bibinfo {author} {\bibfnamefont {K.}~\bibnamefont
  {Nakamura}}, \bibinfo {author} {\bibfnamefont {T.}~\bibnamefont {Akiyama}},
  \bibinfo {author} {\bibfnamefont {T.}~\bibnamefont {Ito}}, \bibinfo {author}
  {\bibfnamefont {M.}~\bibnamefont {Weinert}}, \ and\ \bibinfo {author}
  {\bibfnamefont {A.~J.}\ \bibnamefont {Freeman}},\ }\href {\doibase
  10.1103/PhysRevB.81.220409} {\bibfield  {journal} {\bibinfo  {journal}
  {Physical Review B}\ }\textbf {\bibinfo {volume} {81}},\ \bibinfo {pages}
  {220409} (\bibinfo {year} {2010})}\BibitemShut {NoStop}%
\bibitem [{\citenamefont {Niranjan}\ \emph {et~al.}(2010)\citenamefont
  {Niranjan}, \citenamefont {Duan}, \citenamefont {Jaswal},\ and\ \citenamefont
  {Tsymbal}}]{niranjan2010}%
  \BibitemOpen
  \bibfield  {author} {\bibinfo {author} {\bibfnamefont {M.~K.}\ \bibnamefont
  {Niranjan}}, \bibinfo {author} {\bibfnamefont {C.-G.}\ \bibnamefont {Duan}},
  \bibinfo {author} {\bibfnamefont {S.~S.}\ \bibnamefont {Jaswal}}, \ and\
  \bibinfo {author} {\bibfnamefont {E.~Y.}\ \bibnamefont {Tsymbal}},\ }\href
  {\doibase 10.1063/1.3443658} {\bibfield  {journal} {\bibinfo  {journal}
  {Applied Physics Letters}\ }\textbf {\bibinfo {volume} {96}},\ \bibinfo
  {pages} {222504} (\bibinfo {year} {2010})}\BibitemShut {NoStop}%
\bibitem [{\citenamefont {Yoshikawa}\ \emph {et~al.}(2014)\citenamefont
  {Yoshikawa}, \citenamefont {Obata}, \citenamefont {Taguchi}, \citenamefont
  {Haraguchi},\ and\ \citenamefont {Oda}}]{yoshikawa2014}%
  \BibitemOpen
  \bibfield  {author} {\bibinfo {author} {\bibfnamefont {D.}~\bibnamefont
  {Yoshikawa}}, \bibinfo {author} {\bibfnamefont {M.}~\bibnamefont {Obata}},
  \bibinfo {author} {\bibfnamefont {Y.}~\bibnamefont {Taguchi}}, \bibinfo
  {author} {\bibfnamefont {S.}~\bibnamefont {Haraguchi}}, \ and\ \bibinfo
  {author} {\bibfnamefont {T.}~\bibnamefont {Oda}},\ }\href {\doibase
  10.7567/APEX.7.113005} {\bibfield  {journal} {\bibinfo  {journal} {Applied
  Physics Express}\ }\textbf {\bibinfo {volume} {7}},\ \bibinfo {pages}
  {113005} (\bibinfo {year} {2014})}\BibitemShut {NoStop}%
\bibitem [{\citenamefont {Miwa}\ \emph {et~al.}(2015)\citenamefont {Miwa},
  \citenamefont {Matsuda}, \citenamefont {Tanaka}, \citenamefont {Kotani},
  \citenamefont {Goto}, \citenamefont {Nakamura},\ and\ \citenamefont
  {Suzuki}}]{miwa2015}%
  \BibitemOpen
  \bibfield  {author} {\bibinfo {author} {\bibfnamefont {S.}~\bibnamefont
  {Miwa}}, \bibinfo {author} {\bibfnamefont {K.}~\bibnamefont {Matsuda}},
  \bibinfo {author} {\bibfnamefont {K.}~\bibnamefont {Tanaka}}, \bibinfo
  {author} {\bibfnamefont {Y.}~\bibnamefont {Kotani}}, \bibinfo {author}
  {\bibfnamefont {M.}~\bibnamefont {Goto}}, \bibinfo {author} {\bibfnamefont
  {T.}~\bibnamefont {Nakamura}}, \ and\ \bibinfo {author} {\bibfnamefont
  {Y.}~\bibnamefont {Suzuki}},\ }\href {\doibase 10.1063/1.4934568} {\bibfield
  {journal} {\bibinfo  {journal} {Applied Physics Letters}\ }\textbf {\bibinfo
  {volume} {107}},\ \bibinfo {pages} {162402} (\bibinfo {year}
  {2015})}\BibitemShut {NoStop}%
\bibitem [{\citenamefont {Nozaki}\ \emph {et~al.}(2016)\citenamefont {Nozaki},
  \citenamefont {Kozioł-Rachwał}, \citenamefont {Skowroński}, \citenamefont
  {Zayets}, \citenamefont {Shiota}, \citenamefont {Tamaru}, \citenamefont
  {Kubota}, \citenamefont {Fukushima}, \citenamefont {Yuasa},\ and\
  \citenamefont {Suzuki}}]{nozaki2016}%
  \BibitemOpen
  \bibfield  {author} {\bibinfo {author} {\bibfnamefont {T.}~\bibnamefont
  {Nozaki}}, \bibinfo {author} {\bibfnamefont {A.}~\bibnamefont
  {Kozioł-Rachwał}}, \bibinfo {author} {\bibfnamefont {W.}~\bibnamefont
  {Skowroński}}, \bibinfo {author} {\bibfnamefont {V.}~\bibnamefont {Zayets}},
  \bibinfo {author} {\bibfnamefont {Y.}~\bibnamefont {Shiota}}, \bibinfo
  {author} {\bibfnamefont {S.}~\bibnamefont {Tamaru}}, \bibinfo {author}
  {\bibfnamefont {H.}~\bibnamefont {Kubota}}, \bibinfo {author} {\bibfnamefont
  {A.}~\bibnamefont {Fukushima}}, \bibinfo {author} {\bibfnamefont
  {S.}~\bibnamefont {Yuasa}}, \ and\ \bibinfo {author} {\bibfnamefont
  {Y.}~\bibnamefont {Suzuki}},\ }\href {\doibase
  10.1103/PhysRevApplied.5.044006} {\bibfield  {journal} {\bibinfo  {journal}
  {Physical Review Applied}\ }\textbf {\bibinfo {volume} {5}},\ \bibinfo
  {pages} {044006} (\bibinfo {year} {2016})}\BibitemShut {NoStop}%
\bibitem [{\citenamefont {Qingyi}\ \emph {et~al.}(2017)\citenamefont {Qingyi},
  \citenamefont {Wen}, \citenamefont {Sukegawa}, \citenamefont {Kasai},
  \citenamefont {Seki}, \citenamefont {Kubota}, \citenamefont {Takanashi},\
  and\ \citenamefont {Mitani}}]{qingyi2017}%
  \BibitemOpen
  \bibfield  {author} {\bibinfo {author} {\bibfnamefont {X.}~\bibnamefont
  {Qingyi}}, \bibinfo {author} {\bibfnamefont {Z.}~\bibnamefont {Wen}},
  \bibinfo {author} {\bibfnamefont {H.}~\bibnamefont {Sukegawa}}, \bibinfo
  {author} {\bibfnamefont {S.}~\bibnamefont {Kasai}}, \bibinfo {author}
  {\bibfnamefont {T.}~\bibnamefont {Seki}}, \bibinfo {author} {\bibfnamefont
  {T.}~\bibnamefont {Kubota}}, \bibinfo {author} {\bibfnamefont
  {K.}~\bibnamefont {Takanashi}}, \ and\ \bibinfo {author} {\bibfnamefont
  {S.}~\bibnamefont {Mitani}},\ }\href {\doibase 10.1088/1361-6463/aa87ab}
  {\bibfield  {journal} {\bibinfo  {journal} {Journal of Physics D: Applied
  Physics}\ } (\bibinfo {year} {2017}),\ 10.1088/1361-6463/aa87ab}\BibitemShut
  {NoStop}%
\bibitem [{\citenamefont {Zhang}\ \emph {et~al.}(2009)\citenamefont {Zhang},
  \citenamefont {Richter}, \citenamefont {Koepernik}, \citenamefont {Opahle},
  \citenamefont {Tasnádi},\ and\ \citenamefont {Eschrig}}]{zhang2009}%
  \BibitemOpen
  \bibfield  {author} {\bibinfo {author} {\bibfnamefont {H.}~\bibnamefont
  {Zhang}}, \bibinfo {author} {\bibfnamefont {M.}~\bibnamefont {Richter}},
  \bibinfo {author} {\bibfnamefont {K.}~\bibnamefont {Koepernik}}, \bibinfo
  {author} {\bibfnamefont {I.}~\bibnamefont {Opahle}}, \bibinfo {author}
  {\bibfnamefont {F.}~\bibnamefont {Tasnádi}}, \ and\ \bibinfo {author}
  {\bibfnamefont {H.}~\bibnamefont {Eschrig}},\ }\href {\doibase
  10.1088/1367-2630/11/4/043007} {\bibfield  {journal} {\bibinfo  {journal}
  {New Journal of Physics}\ }\textbf {\bibinfo {volume} {11}},\ \bibinfo
  {pages} {043007} (\bibinfo {year} {2009})}\BibitemShut {NoStop}%
\bibitem [{\citenamefont {Tsujikawa}\ \emph {et~al.}(2012)\citenamefont
  {Tsujikawa}, \citenamefont {Haraguchi},\ and\ \citenamefont
  {Oda}}]{tsujikawa2012}%
  \BibitemOpen
  \bibfield  {author} {\bibinfo {author} {\bibfnamefont {M.}~\bibnamefont
  {Tsujikawa}}, \bibinfo {author} {\bibfnamefont {S.}~\bibnamefont
  {Haraguchi}}, \ and\ \bibinfo {author} {\bibfnamefont {T.}~\bibnamefont
  {Oda}},\ }\href {\doibase 10.1063/1.3703682} {\bibfield  {journal} {\bibinfo
  {journal} {Journal of Applied Physics}\ }\textbf {\bibinfo {volume} {111}},\
  \bibinfo {pages} {083910} (\bibinfo {year} {2012})}\BibitemShut {NoStop}%
\bibitem [{\citenamefont {Miwa}\ \emph {et~al.}(2017)\citenamefont {Miwa},
  \citenamefont {Suzuki}, \citenamefont {Tsujikawa}, \citenamefont {Matsuda},
  \citenamefont {Nozaki}, \citenamefont {Tanaka}, \citenamefont {Tsukahara},
  \citenamefont {Nawaoka}, \citenamefont {Goto}, \citenamefont {Kotani},
  \citenamefont {Ohkubo}, \citenamefont {Bonell}, \citenamefont {Tamura},
  \citenamefont {Hono}, \citenamefont {Nakamura}, \citenamefont {Shirai},
  \citenamefont {Yuasa},\ and\ \citenamefont {Suzuki}}]{miwa2017}%
  \BibitemOpen
  \bibfield  {author} {\bibinfo {author} {\bibfnamefont {S.}~\bibnamefont
  {Miwa}}, \bibinfo {author} {\bibfnamefont {M.}~\bibnamefont {Suzuki}},
  \bibinfo {author} {\bibfnamefont {M.}~\bibnamefont {Tsujikawa}}, \bibinfo
  {author} {\bibfnamefont {K.}~\bibnamefont {Matsuda}}, \bibinfo {author}
  {\bibfnamefont {T.}~\bibnamefont {Nozaki}}, \bibinfo {author} {\bibfnamefont
  {K.}~\bibnamefont {Tanaka}}, \bibinfo {author} {\bibfnamefont
  {T.}~\bibnamefont {Tsukahara}}, \bibinfo {author} {\bibfnamefont
  {K.}~\bibnamefont {Nawaoka}}, \bibinfo {author} {\bibfnamefont
  {M.}~\bibnamefont {Goto}}, \bibinfo {author} {\bibfnamefont {Y.}~\bibnamefont
  {Kotani}}, \bibinfo {author} {\bibfnamefont {T.}~\bibnamefont {Ohkubo}},
  \bibinfo {author} {\bibfnamefont {F.}~\bibnamefont {Bonell}}, \bibinfo
  {author} {\bibfnamefont {E.}~\bibnamefont {Tamura}}, \bibinfo {author}
  {\bibfnamefont {K.}~\bibnamefont {Hono}}, \bibinfo {author} {\bibfnamefont
  {T.}~\bibnamefont {Nakamura}}, \bibinfo {author} {\bibfnamefont
  {M.}~\bibnamefont {Shirai}}, \bibinfo {author} {\bibfnamefont
  {S.}~\bibnamefont {Yuasa}}, \ and\ \bibinfo {author} {\bibfnamefont
  {Y.}~\bibnamefont {Suzuki}},\ }\href {\doibase 10.1038/ncomms15848}
  {\bibfield  {journal} {\bibinfo  {journal} {Nature Communications}\ }\textbf
  {\bibinfo {volume} {8}},\ \bibinfo {pages} {ncomms15848} (\bibinfo {year}
  {2017})}\BibitemShut {NoStop}%
\bibitem [{\citenamefont {Daalderop}\ \emph {et~al.}(1990)\citenamefont
  {Daalderop}, \citenamefont {Kelly},\ and\ \citenamefont
  {Schuurmans}}]{daalderop_1990}%
  \BibitemOpen
  \bibfield  {author} {\bibinfo {author} {\bibfnamefont {G.~H.~O.}\
  \bibnamefont {Daalderop}}, \bibinfo {author} {\bibfnamefont {P.~J.}\
  \bibnamefont {Kelly}}, \ and\ \bibinfo {author} {\bibfnamefont {M.~F.~H.}\
  \bibnamefont {Schuurmans}},\ }\href {\doibase 10.1103/PhysRevB.42.7270}
  {\bibfield  {journal} {\bibinfo  {journal} {Physical Review B}\ }\textbf
  {\bibinfo {volume} {42}},\ \bibinfo {pages} {7270} (\bibinfo {year}
  {1990})}\BibitemShut {NoStop}%
\bibitem [{\citenamefont {Daalderop}\ \emph {et~al.}(1994)\citenamefont
  {Daalderop}, \citenamefont {Kelly},\ and\ \citenamefont
  {Schuurmans}}]{daalderop_1994}%
  \BibitemOpen
  \bibfield  {author} {\bibinfo {author} {\bibfnamefont {G.~H.~O.}\
  \bibnamefont {Daalderop}}, \bibinfo {author} {\bibfnamefont {P.~J.}\
  \bibnamefont {Kelly}}, \ and\ \bibinfo {author} {\bibfnamefont {M.~F.~H.}\
  \bibnamefont {Schuurmans}},\ }\href {\doibase 10.1103/PhysRevB.50.9989}
  {\bibfield  {journal} {\bibinfo  {journal} {Physical Review B}\ }\textbf
  {\bibinfo {volume} {50}},\ \bibinfo {pages} {9989} (\bibinfo {year}
  {1994})}\BibitemShut {NoStop}%
\bibitem [{\citenamefont {Bruno}(1989)}]{bruno_1989}%
  \BibitemOpen
  \bibfield  {author} {\bibinfo {author} {\bibfnamefont {P.}~\bibnamefont
  {Bruno}},\ }\href {\doibase 10.1103/PhysRevB.39.865} {\bibfield  {journal}
  {\bibinfo  {journal} {Physical Review B}\ }\textbf {\bibinfo {volume} {39}},\
  \bibinfo {pages} {865} (\bibinfo {year} {1989})}\BibitemShut {NoStop}%
\bibitem [{\citenamefont {Wang}\ \emph {et~al.}(1993)\citenamefont {Wang},
  \citenamefont {Wu},\ and\ \citenamefont {Freeman}}]{wang_1993a}%
  \BibitemOpen
  \bibfield  {author} {\bibinfo {author} {\bibfnamefont {D.-s.}\ \bibnamefont
  {Wang}}, \bibinfo {author} {\bibfnamefont {R.}~\bibnamefont {Wu}}, \ and\
  \bibinfo {author} {\bibfnamefont {A.~J.}\ \bibnamefont {Freeman}},\ }\href
  {\doibase 10.1103/PhysRevB.47.14932} {\bibfield  {journal} {\bibinfo
  {journal} {Physical Review B}\ }\textbf {\bibinfo {volume} {47}},\ \bibinfo
  {pages} {14932} (\bibinfo {year} {1993})}\BibitemShut {NoStop}%
\bibitem [{\citenamefont {Gould}\ \emph {et~al.}(2004)\citenamefont {Gould},
  \citenamefont {R{\"u}ster}, \citenamefont {Jungwirth}, \citenamefont
  {Girgis}, \citenamefont {Schott}, \citenamefont {Giraud}, \citenamefont
  {Brunner}, \citenamefont {Schmidt},\ and\ \citenamefont
  {Molenkamp}}]{gould_2004}%
  \BibitemOpen
  \bibfield  {author} {\bibinfo {author} {\bibfnamefont {C.}~\bibnamefont
  {Gould}}, \bibinfo {author} {\bibfnamefont {C.}~\bibnamefont {R{\"u}ster}},
  \bibinfo {author} {\bibfnamefont {T.}~\bibnamefont {Jungwirth}}, \bibinfo
  {author} {\bibfnamefont {E.}~\bibnamefont {Girgis}}, \bibinfo {author}
  {\bibfnamefont {G.~M.}\ \bibnamefont {Schott}}, \bibinfo {author}
  {\bibfnamefont {R.}~\bibnamefont {Giraud}}, \bibinfo {author} {\bibfnamefont
  {K.}~\bibnamefont {Brunner}}, \bibinfo {author} {\bibfnamefont
  {G.}~\bibnamefont {Schmidt}}, \ and\ \bibinfo {author} {\bibfnamefont
  {L.~W.}\ \bibnamefont {Molenkamp}},\ }\href {\doibase
  10.1103/PhysRevLett.93.117203} {\bibfield  {journal} {\bibinfo  {journal}
  {Physical Review Letters}\ }\textbf {\bibinfo {volume} {93}},\ \bibinfo
  {pages} {117203} (\bibinfo {year} {2004})}\BibitemShut {NoStop}%
\bibitem [{\citenamefont {Park}\ \emph {et~al.}(2008)\citenamefont {Park},
  \citenamefont {Wunderlich}, \citenamefont {Williams}, \citenamefont {Joo},
  \citenamefont {Jung}, \citenamefont {Shin}, \citenamefont {Olejn{\'\i}k},
  \citenamefont {Shick},\ and\ \citenamefont {Jungwirth}}]{park_2008}%
  \BibitemOpen
  \bibfield  {author} {\bibinfo {author} {\bibfnamefont {B.~G.}\ \bibnamefont
  {Park}}, \bibinfo {author} {\bibfnamefont {J.}~\bibnamefont {Wunderlich}},
  \bibinfo {author} {\bibfnamefont {D.~A.}\ \bibnamefont {Williams}}, \bibinfo
  {author} {\bibfnamefont {S.~J.}\ \bibnamefont {Joo}}, \bibinfo {author}
  {\bibfnamefont {K.~Y.}\ \bibnamefont {Jung}}, \bibinfo {author}
  {\bibfnamefont {K.~H.}\ \bibnamefont {Shin}}, \bibinfo {author}
  {\bibfnamefont {K.}~\bibnamefont {Olejn{\'\i}k}}, \bibinfo {author}
  {\bibfnamefont {A.~B.}\ \bibnamefont {Shick}}, \ and\ \bibinfo {author}
  {\bibfnamefont {T.}~\bibnamefont {Jungwirth}},\ }\href {\doibase
  10.1103/PhysRevLett.100.087204} {\bibfield  {journal} {\bibinfo  {journal}
  {Physical Review Letters}\ }\textbf {\bibinfo {volume} {100}},\ \bibinfo
  {pages} {087204} (\bibinfo {year} {2008})}\BibitemShut {NoStop}%
\bibitem [{\citenamefont {Saito}\ \emph {et~al.}(2005)\citenamefont {Saito},
  \citenamefont {Yuasa},\ and\ \citenamefont {Ando}}]{saito_2005}%
  \BibitemOpen
  \bibfield  {author} {\bibinfo {author} {\bibfnamefont {H.}~\bibnamefont
  {Saito}}, \bibinfo {author} {\bibfnamefont {S.}~\bibnamefont {Yuasa}}, \ and\
  \bibinfo {author} {\bibfnamefont {K.}~\bibnamefont {Ando}},\ }\href {\doibase
  10.1103/PhysRevLett.95.086604} {\bibfield  {journal} {\bibinfo  {journal}
  {Physical Review Letters}\ }\textbf {\bibinfo {volume} {95}},\ \bibinfo
  {pages} {086604} (\bibinfo {year} {2005})}\BibitemShut {NoStop}%
\bibitem [{\citenamefont {Shick}\ \emph {et~al.}(2006)\citenamefont {Shick},
  \citenamefont {M{\'a}ca}, \citenamefont {Ma{\v s}ek},\ and\ \citenamefont
  {Jungwirth}}]{shick_2006}%
  \BibitemOpen
  \bibfield  {author} {\bibinfo {author} {\bibfnamefont {A.~B.}\ \bibnamefont
  {Shick}}, \bibinfo {author} {\bibfnamefont {F.}~\bibnamefont {M{\'a}ca}},
  \bibinfo {author} {\bibfnamefont {J.}~\bibnamefont {Ma{\v s}ek}}, \ and\
  \bibinfo {author} {\bibfnamefont {T.}~\bibnamefont {Jungwirth}},\ }\href
  {\doibase 10.1103/PhysRevB.73.024418} {\bibfield  {journal} {\bibinfo
  {journal} {Physical Review B}\ }\textbf {\bibinfo {volume} {73}},\ \bibinfo
  {pages} {024418} (\bibinfo {year} {2006})}\BibitemShut {NoStop}%
\bibitem [{\citenamefont {Ikeura}\ \emph {et~al.}(2018)\citenamefont {Ikeura},
  \citenamefont {Nozaki}, \citenamefont {Shiota}, \citenamefont {Yamamoto},
  \citenamefont {Imamura}, \citenamefont {Kubota}, \citenamefont {Fukushima},
  \citenamefont {Suzuki},\ and\ \citenamefont {Yuasa}}]{ikeura_2018}%
  \BibitemOpen
  \bibfield  {author} {\bibinfo {author} {\bibfnamefont {T.}~\bibnamefont
  {Ikeura}}, \bibinfo {author} {\bibfnamefont {T.}~\bibnamefont {Nozaki}},
  \bibinfo {author} {\bibfnamefont {Y.}~\bibnamefont {Shiota}}, \bibinfo
  {author} {\bibfnamefont {T.}~\bibnamefont {Yamamoto}}, \bibinfo {author}
  {\bibfnamefont {H.}~\bibnamefont {Imamura}}, \bibinfo {author} {\bibfnamefont
  {H.}~\bibnamefont {Kubota}}, \bibinfo {author} {\bibfnamefont
  {A.}~\bibnamefont {Fukushima}}, \bibinfo {author} {\bibfnamefont
  {Y.}~\bibnamefont {Suzuki}}, \ and\ \bibinfo {author} {\bibfnamefont
  {S.}~\bibnamefont {Yuasa}},\ }\href {\doibase 10.7567/JJAP.57.040311}
  {\bibfield  {journal} {\bibinfo  {journal} {Japanese Journal of Applied
  Physics}\ }\textbf {\bibinfo {volume} {57}},\ \bibinfo {pages} {040311}
  (\bibinfo {year} {2018})}\BibitemShut {NoStop}%
\bibitem [{\citenamefont {Yoda}\ \emph {et~al.}(2016)\citenamefont {Yoda},
  \citenamefont {Shimomura}, \citenamefont {Ohsawa}, \citenamefont {Shirotori},
  \citenamefont {Kato}, \citenamefont {Inokuchi}, \citenamefont {Kamiguchi},
  \citenamefont {Altansargai}, \citenamefont {Saito}, \citenamefont {Koi},
  \citenamefont {Sugiyama}, \citenamefont {Oikawa}, \citenamefont {Shimizu},
  \citenamefont {Ishikawa}, \citenamefont {Ikegami},\ and\ \citenamefont
  {Kurobe}}]{yoda_2016}%
  \BibitemOpen
  \bibfield  {author} {\bibinfo {author} {\bibfnamefont {H.}~\bibnamefont
  {Yoda}}, \bibinfo {author} {\bibfnamefont {N.}~\bibnamefont {Shimomura}},
  \bibinfo {author} {\bibfnamefont {Y.}~\bibnamefont {Ohsawa}}, \bibinfo
  {author} {\bibfnamefont {S.}~\bibnamefont {Shirotori}}, \bibinfo {author}
  {\bibfnamefont {Y.}~\bibnamefont {Kato}}, \bibinfo {author} {\bibfnamefont
  {T.}~\bibnamefont {Inokuchi}}, \bibinfo {author} {\bibfnamefont
  {Y.}~\bibnamefont {Kamiguchi}}, \bibinfo {author} {\bibfnamefont
  {B.}~\bibnamefont {Altansargai}}, \bibinfo {author} {\bibfnamefont
  {Y.}~\bibnamefont {Saito}}, \bibinfo {author} {\bibfnamefont
  {K.}~\bibnamefont {Koi}}, \bibinfo {author} {\bibfnamefont {H.}~\bibnamefont
  {Sugiyama}}, \bibinfo {author} {\bibfnamefont {S.}~\bibnamefont {Oikawa}},
  \bibinfo {author} {\bibfnamefont {M.}~\bibnamefont {Shimizu}}, \bibinfo
  {author} {\bibfnamefont {M.}~\bibnamefont {Ishikawa}}, \bibinfo {author}
  {\bibfnamefont {K.}~\bibnamefont {Ikegami}}, \ and\ \bibinfo {author}
  {\bibfnamefont {A.}~\bibnamefont {Kurobe}},\ }in\ \href {\doibase
  10.1109/IEDM.2016.7838495} {\emph {\bibinfo {booktitle} {2016 {{IEEE
  International Electron Devices Meeting}} ({{IEDM}})}}}\ (\bibinfo {year}
  {2016})\ pp.\ \bibinfo {pages} {27.6.1--27.6.4}\BibitemShut {NoStop}%
\bibitem [{\citenamefont {Rana}\ and\ \citenamefont {Otani}(2018)}]{rana_2018}%
  \BibitemOpen
  \bibfield  {author} {\bibinfo {author} {\bibfnamefont {B.}~\bibnamefont
  {Rana}}\ and\ \bibinfo {author} {\bibfnamefont {Y.}~\bibnamefont {Otani}},\
  }\href {\doibase 10.1103/PhysRevApplied.9.014033} {\bibfield  {journal}
  {\bibinfo  {journal} {Physical Review Applied}\ }\textbf {\bibinfo {volume}
  {9}},\ \bibinfo {pages} {014033} (\bibinfo {year} {2018})}\BibitemShut
  {NoStop}%
\end{thebibliography}%

\clearpage
\newpage
\onecolumngrid


\end{document}